\documentclass[twocolumn]{emulateapj}

\def\alwaysmath#1{\ifmmode{#1}\else{$#1$}\fi}

\lefthead{Law et al. 2007}
\righthead{SPECTROSCOPY OF HIGH REDSHIFT GALAXIES}

\begin{document}

\newcommand{\msun}{\ensuremath{\rm M_\odot}}
\newcommand{\msunyr}{\ensuremath{\rm M_{\odot}\;{\rm yr}^{-1}}}
\newcommand{\Ha}{\ensuremath{\rm H\alpha}}
\newcommand{\Hb}{\ensuremath{\rm H\beta}}
\newcommand{\lya}{\ensuremath{\rm Ly\alpha}}
\newcommand{\Ntwo}{[\ion{N}{2}]}
\newcommand{\kms}{km~s\ensuremath{^{-1}\,}}
\newcommand{\ztwo}{\ensuremath{z\sim2}}
\newcommand{\zthree}{\ensuremath{z\sim3}}

\title{INTEGRAL FIELD SPECTROSCOPY OF HIGH-REDSHIFT STAR FORMING GALAXIES WITH LASER GUIDED ADAPTIVE OPTICS: EVIDENCE
FOR DISPERSION-DOMINATED KINEMATICS\altaffilmark{1}}
\author{\sc David R. Law\altaffilmark{2}, Charles C. Steidel\altaffilmark{2}, Dawn K. Erb\altaffilmark{3}, James E. Larkin\altaffilmark{4},
Max Pettini\altaffilmark{5}, Alice E. Shapley\altaffilmark{6}, and Shelley A. Wright\altaffilmark{4}}

\altaffiltext{1}{Based on data obtained at the W. M. Keck Observatory, which is operated as a scientific partnership among the California Institute of Technology, the University of California, and NASA, and was made possible by the generous financial support of the W. M. Keck Foundation.}
\altaffiltext{2}{Department of Astronomy, California Institute of Technology, MS 105-24,
Pasadena, CA 91125 (drlaw, ccs@astro.caltech.edu)}
\altaffiltext{3}{Harvard-Smithsonian Center for Astrophysics, MS 20, 60 Garden St, Cambridge, MA 02138 (derb@cfa.harvard.edu)}
\altaffiltext{4}{Department of Physics and Astronomy, University of California, Los Angeles, CA 90095;
larkin,saw@astro.ucla.edu}
\altaffiltext{5}{Institute of Astronomy, Madingley Road, Cambridge CB3 0HA, UK (pettini@ast.cam.ac.uk)}
\altaffiltext{6}{Department of Astrophysical Sciences, Princeton University, Peyton Hall, Ivy Lane, Princeton, NJ 08544 (aes@astro.princeton.edu)}

\begin{abstract}

We present early results from an ongoing study of the kinematic structure of star-forming galaxies at
redshift $z \sim 2 - 3$ using integral-field spectroscopy of rest-frame optical nebular emission lines
in combination with Keck laser guide star adaptive optics (LGSAO).
We show kinematic maps of 3 target galaxies Q1623-BX453, Q0449-BX93, and DSF2237a-C2 located at redshifts $z =  2.1820$, 2.0067, and 3.3172
respectively, each of which
is well-resolved with a PSF measuring approximately 0.11 - 0.15 arcsec ($\sim 900$ - 1200 pc at $z \sim 2-3$) after cosmetic smoothing.
Neither galaxy at $z \sim 2$ exhibits substantial kinematic structure
on scales $\gtrsim 30$ km s$^{-1}$; both are instead consistent with largely dispersion-dominated velocity fields
with $\sigma \sim 80$ km s$^{-1}$ along any given line of sight into the galaxy.
While the primary emission component of Q0449-BX93 shows no spatially-resolved kinematic structure, a faint, kinematically distinct emission
region is superposed on the primary region at a relative velocity of $\sim 180$ km s$^{-1}$, suggesting the possible presence of a merging
satellite galaxy.
In contrast, DSF2237a-C2 presents a well-resolved
gradient in velocity  over a distance of $\sim$ 4 kpc with peak-to-peak amplitude of 140 km s$^{-1}$.
This velocity shear was previously
undetected in seeing-limited long-slit observations  despite serendipitous alignment of the slit with the kinematic major axis,
highlighting the importance of LGSAO for understanding velocity structure on subarcsecond scales.
It is unlikely that DSF2237a-C2 represents a dynamically cold rotating disk of ionized gas as the local velocity dispersion
of the galaxy ($\sigma = 79$ km s$^{-1}$) is
comparable to the observed shear.
Using extant multi-wavelength spectroscopy and photometry we relate these kinematic data to physical properties such as stellar mass, gas fraction,
star formation rate, and outflow kinematics and consider the applicability of current galaxy formation models.
While some gas cooling models reproduce the observed kinematics better than a simple rotating disk model, even these provide a poor
overall description of the target galaxies, suggesting that our current understanding of gas cooling mechanisms in galaxies in the early universe
is (at best) incomplete.

\end{abstract}


\section{INTRODUCTION}

At high redshifts $z \sim 2-3$ galaxies undergo strong and rapid evolution from highly irregular clumps of star-formation
into the familiar Hubble sequence of the local universe
(Giavalisco et al. 1996, Papovich et al. 2005), during which time they are thought to accumulate the majority of their stellar mass 
(Dickinson et al. 2003).
The evolution of structure and ordered motion in galaxies at such redshifts is therefore of considerable interest
in the attempt to understand the process of galaxy formation.
As such, numerous authors (e.g. Steidel et al. 2004, Papovich et al. 2006, Reddy et al. 2006a, Erb et al. 2006b,
and references therein) 
have used deep UV-IR photometry and spectroscopy to ascertain such global
properties as stellar mass, population age, AGN fraction, star formation rate,
and the strength of large-scale gaseous outflows expelled by rapid starbursts.

These studies have greatly increased our understanding of the physical nature of these galaxies, indicating
(for instance) that the $z \sim 2$ universe contains galaxies in a wide variety of evolutionary states from young and actively
star-forming to massive and passively evolving (Glazebrook et al. 2004, Daddi et al. 2005, Papovich et al. 2006, Reddy et al. 2006a, and references therein) 
and suggesting (based on clustering properties)
that typical star-forming galaxies at $z \sim 2$ may evolve into the elliptical galaxy population of the local
universe (Adelberger et al. 2005a).
However, in general it is unknown 
whether the majority of star formation in these galaxies occurs in flattened disk-like systems
(as anticipated by cosmological models, e.g. Mo, Mau, \& White 1998)
or disordered non-equilibrium configurations, and whether this star formation is due primarily to large-scale gravitational
instabilities, tidal shocks induced by major mergers, or some other as-yet undetermined physical
mechanism.  Similarly, it is unknown
whether individual regions of star formation within a given galaxy follow a global abundance pattern or exhibit
strong variations in chemical enrichment.  

These questions are most readily investigated by means of 
optical nebular emission lines (redshifted into the near-IR at $z \gtrsim 1$)
such as H$\alpha$, H$\beta$, [O\,{\sc iii}] $\lambda$5007, and  [N\,{\sc ii}] $\lambda$6584 which
predominantly trace ionized gas in active star forming regions and can provide a kinematic probe of the gravitational
potential.
Such studies using slit spectroscopy (e.g. Erb et al. 2003,2004,2006a; Weiner et al. 2006)
have met with moderate success, enabling the determination of dynamical masses for a large sample of galaxies at $z \gtrsim 1$,
but are complicated by uncertainties resulting from slit placement, possible misalignment with the kinematic major
axis, and the small angular size of the galaxies (typically less than an arcsecond;
see Fig. 2 of Law et al. 2007) relative to the size of the seeing disk.
The twin complications of slit placement and alignment have recently been overcome using seeing-limited integral-field unit (IFU) spectroscopy to obtain
two-dimensional maps of the ionized gas kinematics within these galaxies.
A sample of the largest and brightest galaxies in the rest-frame UV-selected sample has recently been observed by F{\"o}rster Schreiber et al. (2006),
who find (in 9 out of 14 cases) smoothly varying velocity shear
generally consistent with rotation in a flattened disk-like configuration.
However, it is unknown whether atmospheric seeing (typically $\sim$ half the size of the galaxy) masks more complex structure,
and to what extent the prevalence of velocity shear in these sources is a product of the selection criteria of the study, which
(for observational reasons) preferentially select bright, extended sources many of which exhibited large velocity shear 
or dispersion in previous long-slit studies (Erb et al. 2003, 2006a).

With the aid of laser guide star adaptive
optics (LGSAO) technology it is now possible to study high-redshift galaxies on sub-kiloparsec scales, permitting fruitful observation of galaxies
either unresolved or barely resolved in previous ground-based observations.
Wright et al. (2007) have demonstrated the utility of LGSAO in combination with integral-field spectroscopy for a galaxy at redshift $z \sim 1.5$;
in this contribution we present early results from an ongoing study of galaxies in the redshift range $z \sim 2 - 3$.
In \S 2 we describe our observational program and basic data reduction techniques.  In \S 3 and 4 we describe the morphological properties, globally-integrated
spectra, and line ratio diagnostics of these galaxies as derived from the integral-field data.  In \S 5 we use extant broadband photometry and UV spectroscopy
of the target galaxies to model the stellar populations and baryonic masses of the galaxies.  We present 
spatially resolved maps of the ionized gas kinematics in \S 6, discussing the velocity field of each of our targets
in the context of theoretical models and the physical properties given in \S 3-5.
Finally, in \S 7 we compare our results to those of F{\"o}rster Schreiber et al. (2006) and other groups and discuss what general conclusions can
be drawn about galaxy formation in the early universe.

We assume a standard $\Lambda$CDM cosmology in which $H_0 = 71$ km s$^{-1}$ Mpc$^{-1}$, $\Omega_{\rm m} = 0.27$, 
and $\Omega_{\rm \Lambda} = 0.73$.


\section{OBSERVING AND DATA REDUCTION}

\subsection{Sample Selection and Observational Strategy}

Observations were performed using the OSIRIS (OH Suppressing InfraRed Imaging Spectrograph; Larkin et al. 2006) 
integral-field spectrograph in conjunction with the Keck II 
LGSAO system (Wizinowich et al. 2006, van Dam et al. 2006).  As described
in detail by Larkin et al. (2006), OSIRIS is a lenslet-type spectrograph with a 2048 x 2048 Hawaii II detector array,
spectral resolution $R \sim 3600$, and a complement of reimaging optics which permit sampling of the field
with spectral pixels (spaxels) ranging in angular size from 20 --- 100 milliarcseconds (mas).

We select star forming galaxies at redshifts $z \sim 2 - 3$ (for which H$\alpha$ and [O\,{\sc iii}] $\lambda$5007, hereafter simply [O\,{\sc iii}],
respectively fall into the observed-frame $K$-band) from the spectroscopic redshift
catalog of Steidel et al. (2003, 2004).
Based on the IFU simulations of Law et al. (2006) which explore the flux requirements necessary
for reasonable detection within a few hours of integration using OSIRIS,
we focus on those galaxies which are known to have either nebular line fluxes greater than $\sim$ 10$^{-16}$ erg s$^{-1}$ cm$^{-2}$
based on previous long-slit observations (Erb et al. 2003, 2006a) or
high star formation rates (SFR) as estimated from rest-frame UV photometry (for one galaxy [Q0449-BX93] not targeted with long-slit spectroscopy).
Physical constraints also dictate that we preferentially select galaxies which are within 60 arcseconds of a suitably bright ($R \lesssim 17$ mag) tip-tilt reference
star (as required by the LGSAO system), and which lie at spectroscopic redshifts such that 
nebular emission lines fall between the strong spectroscopically unresolved night-sky OH emission features which dominate the near-IR background.

At $K$-band wavelengths between the OH-emission features, thermal radiation from warm optical components in the light path
is the dominant contributor to the total background flux (see Law et al. 2006).  
Since successful observation of our target galaxies requires performance at the limit of OSIRIS' capability,
we use the second-largest spaxel scale\footnote{Note that 50 mas is approximately the Keck diffraction
limit at $\lambda = 2\micron$.} of 50 mas lenslet$^{-1}$, for which the balance between thermal background and detector noise
is optimal for detection of extremely faint objects.
Using narrow-band order sorting filters ($\Delta\lambda \sim$ 100 nm), 
the OSIRIS field of view in this configuration is approximately 2 $\times$ 3 arcseconds.

Observations were performed on UT 2006 June 5 and 2006 October 4-5; the June data were obtained under exceptional
photometric conditions with visual-band seeing estimates varying between 0.4 and 0.6'' while the October data were obtained in moderate
conditions with seeing varying between 0.7 and 1.0'' over the duration of the observations.
In Table \ref{targets.table} we list the three targets discussed herein as part of our pilot program.
Each target was observed with the aid of the Keck LGSAO system, typically using a magnitude $R \sim 15-16$ tip-tilt (TT) reference star with
an angular separation from the science target of less than an arcminute for a correction strehl of greater than
30\% at $\lambda = 2 \micron$ and an approximately circular PSF with FWHM $\sim 75$/125 mas for the June/October data respectively.
\footnote{The 75 mas measurement is strictly an upper limit since this PSF is undersampled using 50 mas spaxels.  Since science observations are
performed with the TT star off-axis the PSF of the science observations will not perfectly match that of the TT star; however, this generally
does not greatly increase the effective PSF (see Liu et al. 2006).}

A typical observing sequence was established as follows.  First, we obtained short ($\sim$ 1 minute) 
observations of the TT reference star for a given science
target to refine the pointing model, provide an estimate of the PSF, and serve as an approximate check for later flux
calibration.  The telescope was then
blind-offset to the science target using relative offsets calculated from deep ground based optical imaging
of the target fields.
The position angle of OSIRIS for these observations was chosen to ensure that the tip-tilt reference star fell within the unvignetted field
of view of the LGSAO system.
Each target was observed in sets of two 15 
minute\footnote{Due to the read noise characteristics of OSIRIS, long individual exposures
are desirable in order to optimize the S/N ratio of the observations.} 
integrations with the target galaxy alternately
placed 0.7 arcseconds above and below the center of the IFU (positions A and B).  Small dithers
were introduced around these positions on each successive set of exposures.
This 30 minute exposure sequence was repeated for $\sim$ 2-4 hours as necessary to achieve a high-quality detection;
typically it was possible to verify detection of target galaxies in the difference of two 15 minute exposures.

\begin{deluxetable*}{lcccccccc}
\tablecolumns{9}
\tablewidth{0pc}
\tabletypesize{\scriptsize}
\tablecaption{General Information}
\tablehead{
\colhead{Name} & \colhead{RA (J2000)} & \colhead{DEC (J2000)} & \colhead{Observed} & \colhead{Time\tablenotemark{a}} & \colhead{$R_{\rm TT}$\tablenotemark{b}} & \colhead{$\theta_{\rm TT}$}\tablenotemark{c} & \colhead{$\theta_{\rm seeing}$}\tablenotemark{d} & \colhead{$\theta_{\rm PSF}$}\tablenotemark{e}}
\startdata
Q1623-BX453 & 16:25:50.854 & +26:49:31.28 & Jun '06 & 2.5 & 15.1 & 32 & 0.4 & 75/110\\
Q0449-BX93 & 04:52:15.417 & -16:40:56.88 & Oct '06 & 4.5 & 15.8 & 51 & 0.6 & 125/150\\
DSF2237a-C2 & 22:40:08.298 & +11:49:04.89 & Jun '06 & 1.5 & 16.0 & 24 & 0.4 & 75/110\\
\enddata
\tablenotetext{a}{Total observing time in hours.}
\tablenotetext{b}{$R$-band magnitude of TT star for LGSAO correction.}
\tablenotetext{c}{Angular separation (arcseconds) of TT star from target.}
\tablenotetext{d}{Average $K$-band seeing (arcseconds) during observation.}
\tablenotetext{e}{FWHM of the $K$-band PSF (mas) during on-axis TT star observation (before/after spatial smoothing respectively).}
\label{targets.table}
\end{deluxetable*}

\subsection{Reducing IFU Data}

Data reduction was performed using a multi-step process consisting of both Keck/OSIRIS data reduction pipeline (DRP) routines
and custom IDL scripts tailored for use with faint emission line spectra.
Since OSIRIS is a relatively new scientific instrument and standard data reduction methods have yet to
be generally accepted, we present our current data reduction method in detail.

We divide the exposures into two sets taken at positions A and B, and use DRP routines to median combine each set to produce
a sky frame.
These sky frames are then subtracted from each of the respective object frames (i.e. the sky produced by stacking
observations at position B is subtracted from the observations at position A and vice versa).
After sky subtraction, individual exposures are   
extracted from the raw two-dimensional format into three-dimensional data cubes.
This DRP extraction algorithm is complicated and non-intuitive since spectra from individual spaxels physically overlap
on the detector; we refer the reader to the OSIRIS user manual for further information.
Due to the highly time-variable nature of the IR background the residual flux pedestal in these sky-subtracted data cubes
is often unsatisfactorily high and we 
therefore perform a second-pass background subtraction on each of these cubes as follows.
Using custom IDL routines we calculate the median pixel value in each spectral channel
and subtract this value from all pixels within the channel to ensure a zero-flux median in all spectral slices throughout the data cube.
This effectively removes flux pedestals resulting from variable sky brightness, although
small variations in the background persist across the spatial dimensions of each cube.
As OSIRIS data reduction algorithms are progressively refined it may prove possible to further suppress background fluctuations,
achieving effectively deeper integrations.

These sky-subtracted data cubes are spatially registered using the telescope steering system (TSS) offset coordinates contained within
the image headers
and combined using a 3$\sigma$ clipped mean algorithm.
The resulting cube is spatially oversampled by a factor of two in each dimension (i.e. mapped to pixels 25 mas in size), 
rotated with flux-conserving resampling to a standard (North up, East left) orientation, 
and the image slice at each spectral channel convolved with a spatial gaussian kernel of
FWHM 3.2 oversampled pixels (80 mas, giving an effective PSF FWHM of 110/150 mas for June/October data respectively) 
in order to increase the S/N ratio of each spectrum.
Wavelength calibration is automatically performed by the OSIRIS DRP;
we determine that this calibration is generally accurate to within $\sim 0.1\AA$ based on
fits to the centroids of strong OH sky lines 
with regard to fiducial vacuum wavelengths given by Rousselot et al. (2000).  Finally, all spectra are
converted to the heliocentric rest frame using the relative motion solution given by the IRAF/{\it rvcorrect} package.
Based on the measured FWHM of the (unresolved) OH lines, we conclude that the effective spectral resolution of
our data is $R \sim 3600$.\footnote{This represents the narrowest OH line observed; many appear to be broader because the OH forest
is typically composed of unresolved doublets separated by less than $\sim 1$\AA.}

Since we do not dedicate time in each observing sequence to obtaining a clean ``sky'' observation, negative residuals
of the target galaxy are located about 1.4'' above and below the positive image in the final stacked data cube, effectively halving our useful field of view.
However, given the small size of the target
galaxies (UV morphologies indicate that $\sim$ 90\% are less than 1'' in extent; Law et al. 2007) 
we find this approach preferable for maximization of on-source integration time by eliminating the need to frequently
nod off-target
in order to adequately characterize the background flux (although c.f. discussion by Davies 2007).


\section{NEBULAR MORPHOLOGIES}

\begin{figure*}
\plotone{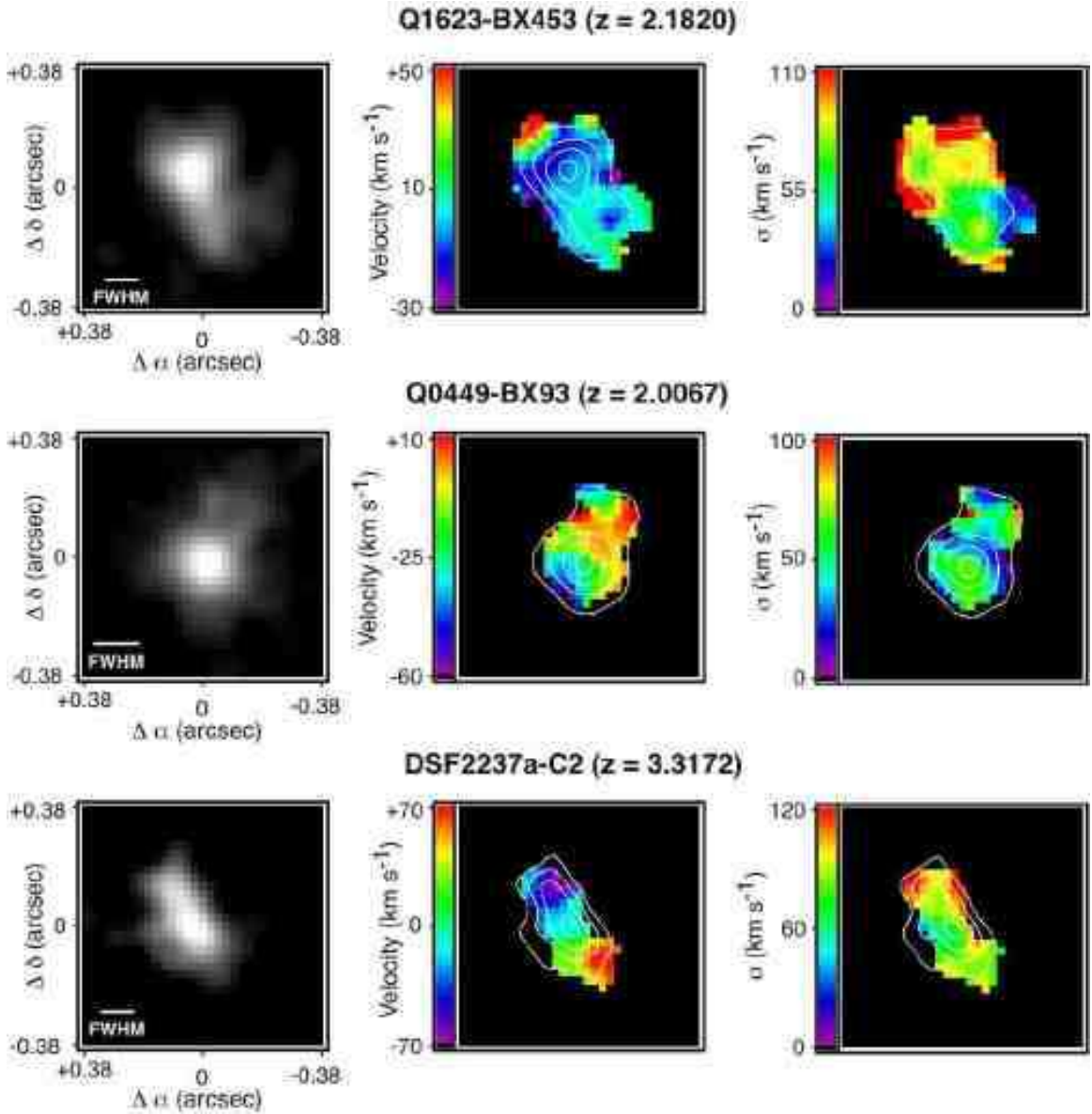}
\caption{
OSIRIS maps of (left to right) nebular emission line flux density (H$\alpha$ for Q1623-BX453 and Q0449-BX93, [O {\sc iii}] for DSF2237a-C2), 
velocity, and velocity dispersion for the three target galaxies.  Stretch on plots of emission
line flux density is linear and runs from zero to the peak flux density of each galaxy, linear contours in flux density are overplotted
on the velocity and dispersion maps.  The FWHM of the PSF after smoothing is 110 mas in Q1623-BX453 and DSF2237a-C2 and 150 mas in Q0449-BX93, 
indicated by solid lines in the left-hand panels.
Individual pixels measure 25 mas, the total field of view (0.75 $\times$ 0.75 arcseconds) is comparable 
to the seeing-limited spatial resolution of previous work.
This field of view corresponds to 6.3, 6.4, and 5.7 physical kpc at the redshift of Q1623-BX453, Q0449-BX93, and DSF2237a-C2 respectively.}
\label{resultsfig.fig}
\end{figure*}

We produce maps of the ionized gas morphology of the three target galaxies by 
collapsing each data cube about the 4-5 spectral channels corresponding to the peak emission wavelength
of the spectral line (H$\alpha$ for Q1623-BX453 and Q0449-BX93, [O\,{\sc iii}] for DSF2237a-C2).
As illustrated in Figure \ref{resultsfig.fig} (top left panel), 
Q1623-BX453 consists of an H$\alpha$ bright nucleation\footnote{We adopt the term ``nucleation'' to qualitatively describe a
concentrated region of high surface brightness that might naively be described as the ``nucleus'' of a given galaxy (see discussion in Law et al. 2007).}
with a lower surface brightness region extending towards the southwest and an extremely faint feature on the western edge.
This faint feature is only barely detected and may be an artifact of imperfect sky subtraction.
While fairly centrally concentrated, the bright emission region is inconsistent with a point source, having a diameter of 200 mas (1.7 kpc) at the half peak-flux
contour compared to (an upper limit of) 110 mas for the PSF of the tip-tilt reference star.
Q0449-BX93 (Fig. \ref{resultsfig.fig}; middle left panel) has a morphology quite similar to that of Q1623-BX453, consisting of
a bright nucleated region with low surface brightness emission extending towards the northwest.  Again, the
bright region is inconsistent with a point source, having diameter 180 mas (1.5 kpc) at the half peak-flux contour.
In contrast, DSF2237a-C2 (Fig. \ref{resultsfig.fig}; bottom left panel) has a more elongated morphology than either Q1623-BX453 or Q0449-BX93, 
with flux distributed fairly evenly along the major axis.

Given the elongated morphology of DSF2237a-C2, one traditional assumption would
be that emission in this galaxy is predominantly from an inclined disk of ionized gas, and that the apparent
ellipticity is due to perspective foreshortening of a circularly symmetric disk.
Under such an assumption, we fit a 2D gaussian to the H$\alpha$ surface brightness map and conclude that with a major/minor axis
ratio of $r_{\rm maj}/r_{\rm min} = 1.53$, such a disk has a position angle of $\theta = 38^{\circ}$ East of North and
an inclination of $i = 49^{\circ}$ away from the normal.

While it is tempting to draw such physical inferences from the ionized-gas morphologies,
these inferences are not necessarily robust given the irregularity of typical star-forming galaxies at redshift $z \sim 2-3$,
which typically consist of 1-3 nucleations connected by
regions of low surface-brightness emission (as viewed in the rest-frame UV; see Fig. 2 of Law et al. 2007).
Since the UV and ionized gas morphologies are strongly correlated (at least in the local universe; Gordon et al. 2004, 
Kennicutt et al. 2004, Lee et al. 2004, although c.f. Conselice et al. 2000)
and trace the brightest regions of star formation unobscured by dust
they are therefore a complex product of the strength and distribution of star forming regions,
dust, viewing angle, and possible merger-induced irregularities, rendering them unreliable indicators
of the galactic potential.
Indeed, recent studies have found little to no correlation between morphology and major-axis velocity shear (Erb et al. 2004) or any of an
assortment of other parameters such as stellar mass, star formation rate, or the strength of galactic-scale outflows (Law et al. 2007).
Even if the rotating gas disk hypothesis were correct, it is not obvious that estimates of the disk size and/or orientation based on the
length of the major/minor axes are meaningful since the observed morphology of such a disk will be dominated by the clumpy distribution
of star formation (see, e.g., theoretical models by Noguchi 1999 and Immeli et al. 2004b).

We quantify an average linear dimension of
the nebular line emission region by adopting a segmentation map which includes
all spaxels (i.e. lenslets) for which nebular emission can be fit with a gaussian profile with S/N ratio (SNR) $\sim$ 6, and which
have a velocity of less than 300 km s$^{-1}$ relative to systemic (see \S 6 for further explanation of these selection criteria).
The effective size of this region is determined as follows.
Assuming that the $N$ elements in the segmentation map are distributed in an approximately circular arrangement, the radius of this circle is given
simply by $r' = \sqrt{N/\pi}$.  The corrected radius $r$ is then determined by subtracting off the radius of the stellar PSF
in quadrature (Table \ref{targets.table}; this is typically 
a small correction of less than $\sim$10\%) and converting to physical kpc at the redshift of the source.  The PSF-corrected emission area is 
defined as $I = \pi r^2$.  
Adopting the non-parametric gini-multiplicity ($G-\Psi$) classification scheme described by Law et al. (2007), we note that the ionized-gas morphologies
of our three target galaxies (Table \ref{OSIRISmorphs.table}) closely resemble the typical rest-frame UV morphologies of $z \sim 2-3$ star forming galaxies
in the GOODS-N (Law et al. 2007) and Q1700 (Peters et al. 2007) fields as seen by HST-ACS, albeit with limiting star formation rate density roughly five
times higher than that of the HST data.

\begin{deluxetable}{lcccc}
\tablecolumns{5}
\tablewidth{0pc}
\tabletypesize{\scriptsize}
\tablecaption{OSIRIS Morphologies.}
\tablehead{
\colhead{Galaxy} & \colhead{$I$\tablenotemark{a}} & \colhead{$r$\tablenotemark{b}} &
\colhead{$G$\tablenotemark{c}} & \colhead{$\Psi$\tablenotemark{d}}}
\startdata
Q1623-BX453 & 9.5 $\pm$ 0.3 & 1.74 $\pm$ 0.03 & 0.33 & 1.48\\
Q0449-BX93 & 4.7 $\pm$ 0.6 & 1.23 $\pm$ 0.07 & 0.25 & 1.32\\
DSF2237a-C2 & 3.3 $\pm$ 0.3 & 1.03 $\pm$ 0.04 & 0.29 & 3.33\\
\enddata
\label{OSIRISmorphs.table}
\tablenotetext{a}{Area of nebular emission (kpc$^2$).  Uncertainty represents half the PSF correction.}
\tablenotetext{b}{Radius of nebular emission (kpc).  Uncertainty represents half the PSF correction.}
\tablenotetext{c}{Gini.}
\tablenotetext{d}{Multiplicity.}
\end{deluxetable}

\section{Global spectra}

\subsection{Flux Calibration}

\begin{figure*}
\plotone{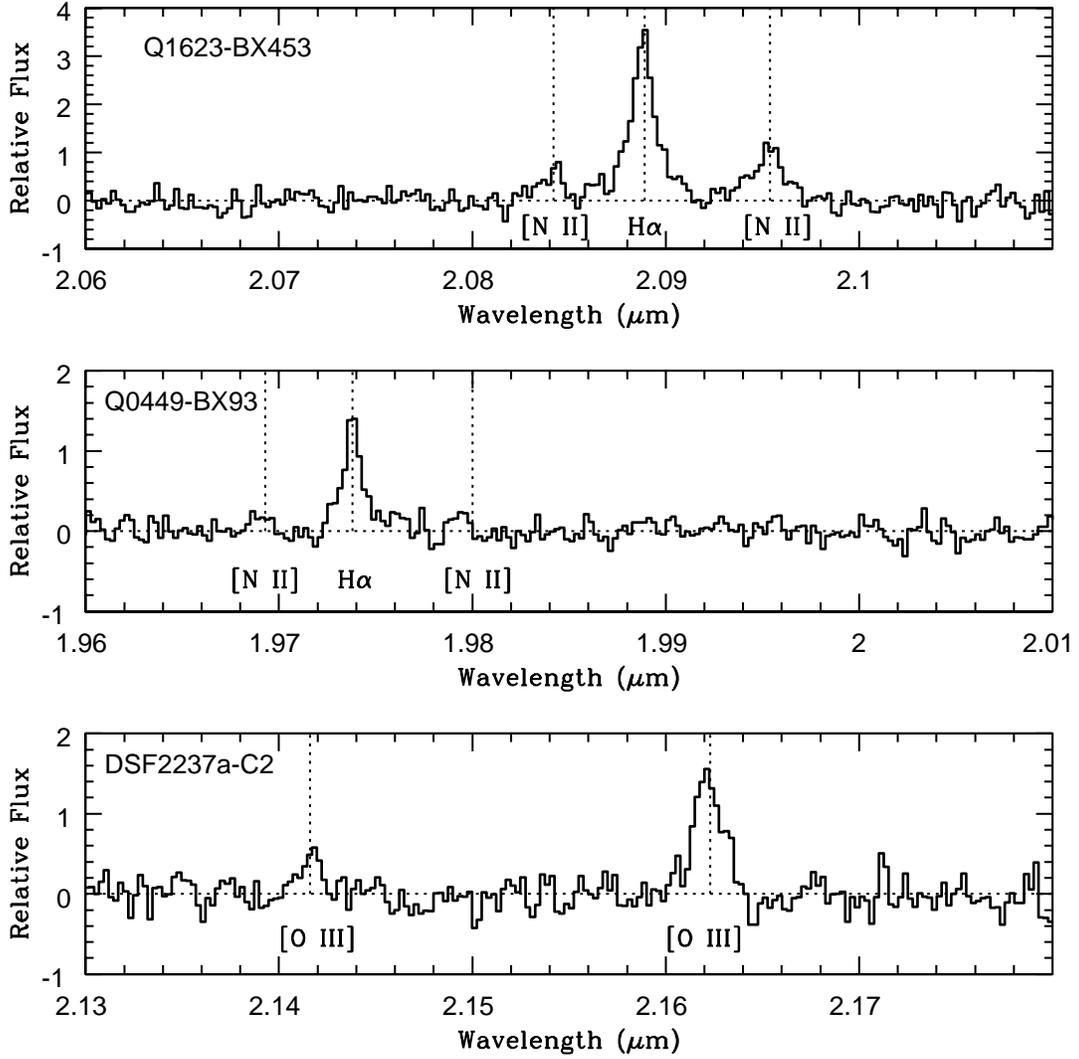}
\caption{OSIRIS spectra of Q1623-BX453, Q0449-BX93, and DSF2237a-C2 collapsed over the spatial extent of each galaxy.
Vertical dotted lines denote the wavelengths of
[N {\sc ii}] $\lambda 6550$, H$\alpha$, and [N {\sc ii}] $\lambda 6585$ emission (for Q1623-BX453 and Q0449-BX 93)
and [O {\sc iii}] $\lambda 4960$ and $\lambda 5008$ emission (for DSF2237a-C2).  Lines for [N {\sc ii}] and [O {\sc iii}] $\lambda 4960$ are drawn at the
fiducial wavelengths expected based on the H$\alpha$ and [O {\sc iii}] $\lambda 5008$ emission redshifts.}
\label{spectra.fig}
\end{figure*}

Integrated spectra (shown in Fig. \ref{spectra.fig}) were generated for each source by 
convolving the data cube during reduction with a broad 150 mas spatial
gaussian kernel, which considerably improves the S/N ratio (SNR) of each spaxel at the cost of suppressing spatial information.  All spaxels
with SNR $>$ 6 were combined to produce a ``clean'' integrated spectrum, while all spaxels within a radius of about 2$r$ were combined to produce
a ``total flux'' integrated spectrum.  The first of these is of higher quality and is used to determine the relative flux in individual emission
lines, while the second spectrum better samples the faint flux distribution and is used to fix the absolute flux calibration.
We eliminate any residual gradients as a function of wavelength by subtracting off a first-order spline fit to the integrated spectrum
using standard IRAF techniques (with iterative rejection to exclude emission lines from the spline fit).
We estimate the flux calibration of our integrated emission-line spectrum using
detailed modelling of the OSIRIS + Keck LGSAO system + atmospheric throughput as a function of wavelength
based on the simulations of Law et al. (2006) and refined to accomodate recent measurements of system throughput parameters.
The results of this wavelength-dependent calibration method agree to within $\sim 25$\% with those obtained using the broadband photometric
zero-point calibrations published in the OSIRIS manual (roughly as close an agreement as may reasonably be
expected since these zero-points average the transmission
function throughout the bandpass).
The greatest uncertainty in flux calibration arises however in the additional requirement that we correct
for the unknown amount of flux lost to the seeing halo of the AO-corrected PSF.  For a point source with
a 35\% strehl correction approximately 35\% of the total light is contained within the diffraction-limited core of FWHM $\sim 50$ mas while the remaining 65\%
is spread throughout a seeing-limited halo.  The percentage of total flux measured for a given observation is therefore
a complex function of the strehl of the AO correction (which depends in turn upon the magnitude/separation of the TT star and the wavelength of observation), 
the quality of the seeing, and the physical structure of the source.
Based on simulations adopting quantities for these various parameters roughly representative of our observing conditions, we expect to recover $\sim 60$\% of
the light from a given target source.
This percentage is roughly consistent with that estimated using measurements of the average flux density in
narrow-band observations of our TT stars compared to $K_s$ magnitudes drawn from the 2MASS database.

Given all of the uncertainties inherent in this bootstrapped flux calibration, we estimate that the systematic flux
uncertainty for a given source is $\sim$ 30\%.
In contrast, the relative flux uncertainty between two emission lines (i.e. the uncertainty in the line flux ratio)
is considerably lower and dominated simply by the root-mean square (RMS) deviation of the integrated spectrum.
Formally, we adopt as the 1$\sigma$ uncertainty in each emission line flux the error due to RMS fluctuations per spectral channel
compounded over the measured width of the emission line.  Similar estimates of the RMS uncertainty may be used to place $3\sigma$
upper limits on the strength of undetected emission lines using a reasonable guess at the likely intrinsic width
of the undetected line.

Based on these calibrations and observations of the three target galaxies we determine that
$f_{\rm lim} = 3 \times 10^{-16}$ erg s$^{-1}$ cm$^{-2}$ arcsec$^{-2}$ 
represents the 6$\sigma$ H$\alpha$ surface brightness threshhold for our observations
(i.e. the faintest believable feature for two hours of integration at $\lambda \sim 2\mu$).
At redshift $z = 2.5$ this is equivalent to a 
detection threshhold of SFR$_{\rm lim} = 1 M_{\odot}$ yr$^{-1}$ kpc$^{-2}$ (adopting the H$\alpha$ flux to SFR calibration
described in \S 5.2).\footnote{Note that this is the threshhold for \textit{detection}, the threshhold for deriving reliable kinematic information is 
approximately 4 $M_{\odot}$ yr$^{-1}$ kpc$^{-2}$.  For comparison, the limiting star formation rate density in the rest-UV morphological analysis presented
by Law et al. (2007) is $\sim$ 0.2 $M_{\odot}$ yr$^{-1}$ kpc$^{-2}$.}
Given that a star formation rate density of $\sim$ 0.1 $M_{\odot}$ yr$^{-1}$ kpc$^{-2}$ represents the threshhold between starburst galaxies
and the centers of normal galactic disks in the local universe (Kennicutt 1998), we do not expect to detect local-type disk galaxies
in the course of our study even if such galaxies existed at high redshift.

In Table \ref{globals.table} we compile the systemic, UV absorption-line, and Ly$\alpha$ emission redshifts and nebular line fluxes
of the target galaxies.
In both cases for which previous NIRSPEC observations were obtained (Q1623-BX453 and DSF2237a-C2) the systemic redshifts
agree to within $\Delta z = 0.0008$ or about 50 km s$^{-1}$ (i.e. consistent to within roughly one fifth the velocity resolution
of NIRSPEC [$R \sim 1400$]).
The integrated H$\alpha$ flux of Q1623-BX453 is 1.7 times greater in the OSIRIS data than the NIRSPEC, consistent with the approximately 50\%
slit-loss
correction factor estimated by Erb et al. (2006a).  However, the  [O\,{\sc iii}]
flux of DSF2237a-C2 appears to be a factor of $\sim$ 2 weaker in the OSIRIS data than the NIRSPEC.
The first, and perhaps most likely, explanation for this discrepancy is that the NIRSPEC flux observed is strongly sensitive to
differential slit losses between the standard star and target galaxy observations, which can drive flux errors in both directions depending on
atmospheric conditions and the quality of the background subtraction in each case.  
However, it is also possible that this galaxy contains a large fraction of its total
flux in the form of extended low surface brightness emission undetected with OSIRIS but detected in the more sensitive long-slit observations.

\begin{deluxetable*}{lccccccccccc}
\tablecolumns{10}
\tablewidth{0pc}
\tabletypesize{\scriptsize}
\tablecaption{Primary Spectroscopic Characteristics}
\tablehead{
\colhead{Galaxy} & \colhead{$\lambda_{\rm neb}$\tablenotemark{a}}  &
\colhead{$z_{\rm neb}$\tablenotemark{b}} & \colhead{$z_{\rm abs}$\tablenotemark{c}} & \colhead{$z_{\rm Ly\alpha}$\tablenotemark{d}} & \colhead{$F_{\rm
    H\alpha}$\tablenotemark{e}} & \colhead{$L_{\rm H\alpha}$\tablenotemark{f}} &
   \colhead{$F_{\rm [O\,{\sc III}]}$\tablenotemark{e}} & 
   \colhead{$L_{\rm [O\,{\sc III}]}$\tablenotemark{f}} &
  \colhead{$\sigma_{\rm
    v}$\tablenotemark{g}} & \colhead{$v_{\rm c}$\tablenotemark{h}} & \colhead{$M_{\rm dyn}$\tablenotemark{i}}} 
\startdata
Q1623-BX453 & 20888.30 & 2.1820 & 2.1724 & 2.1838 & $23.1\pm0.5$ & $18.0 \pm 0.4$  & ... & ... & 92 $\pm$ 8 & ...& 17 $\pm$ 2 \\
Q0449-BX93 & 19737.80 & 2.0067 & 2.004 & ... & $9.3\pm0.3$ & $4.1 \pm 0.1$  & ... & ... & 72 $\pm$ 12 & ... & 7 $\pm$ 2\\
DSF2237a-C2 & 21621.80 & 3.3172 & 3.319 & 3.333 & ... & ... & $10.1\pm 0.5$ & $2.2 \pm 0.1$ & 101/79 $\pm$ 15 & 70 $\pm$ 10 & 5 $\pm$ 1\\
\enddata
\label{globals.table}
\tablenotetext{a}{Vacuum wavelength of peak emission in Angstroms.}
\tablenotetext{b}{Vacuum redshift of nebular emission line: H$\alpha$ for Q1623-BX453 and Q0449-BX93, [O\,{\sc iii}] $\lambda$5007 for DSF2237a-C2.}
\tablenotetext{c}{Vacuum heliocentric redshift of rest-frame UV interstellar absorption lines.}
\tablenotetext{d}{Vacuum heliocentric redshift of Ly$\alpha$ emission.}
\tablenotetext{e}{Emission line flux in units of $10^{-17}$ erg s$^{-1}$ cm$^{-2}$.  Uncertainties quoted are based on random errors,
global systematic uncertainty is $\sim 30$\%.}
\tablenotetext{f}{Extinction-corrected emission line luminosity in units of $10^{42}$ erg s$^{-1}$.  Uncertainties quoted are based on random errors,
global systematic uncertainty is $\sim 30$\%.}
\tablenotetext{g}{Velocity dispersion in units of km s$^{-1}$. Two values are given for DSF2237a-C2, the overall dispersion of the integrated
spectrum and the median velocity dispersion in a given spaxel.}
\tablenotetext{h}{Circular velocity in units of km s$^{-1}$.}
\tablenotetext{i}{Dynamical mass in units of $10^{9} M_{\odot}$.}
\end{deluxetable*}

\begin{deluxetable}{lccc}
\tablecolumns{4}
\tablewidth{0pc}
\tabletypesize{\scriptsize}
\tablecaption{Secondary Spectroscopic Characteristics.}
\tablehead{
\colhead{Galaxy} & \colhead{$F_{\rm [O\,{\sc III}]}
  (\lambda 4960)$\tablenotemark{a}} & \colhead{$F_{\rm [N\,{\sc II}]}
  (\lambda 6549)$\tablenotemark{a}} & \colhead{$F_{\rm [N\,{\sc II}]}
  (\lambda 6585)$\tablenotemark{a}}} 
\startdata
Q1623-BX453 &  ... & $3.3\pm0.4$ & $7.7\pm0.5$\\
Q0449-BX93 &  ... & $<1.1$ & $<1.1$\\
DSF2237a-C2 & $2.5\pm0.5$ & ... & ...\\
\enddata
\label{globalsOSIRIS2.table}
\tablenotetext{a}{Emission line flux in units of $10^{-17}$ erg s$^{-1}$ cm$^{-2}$.  Limits are 3$\sigma$ limits.  Uncertainties quoted are stochastic,
global systematic uncertainty is $\sim 30$\%.}
\end{deluxetable}


\subsection{Global Chemistry}

In the case of Q1623-BX453, both 
[N\,{\sc ii}] $\lambda 6549$ and $\lambda 6584$ (the stronger $\lambda 6584$ line is hereafter simply [N\,{\sc ii}]) are
detected reliably in the integrated galaxy spectrum (see Figure \ref{spectra.fig}), allowing us to estimate the average metallicity
of the galaxy.
Defining the flux ratio $N2 =$ log([N\,{\sc ii}]/H$\alpha$), we follow Pettini \& Pagel (2004) who estimate the
oxygen abundance as
\begin{equation}
12 + \textrm{log}(O/H) = 8.90 + 0.57 \times N2
\label{metallicity.eqn}
\end{equation}

Adopting the [N\,{\sc ii}] flux tabulated in Table \ref{globalsOSIRIS2.table} we find
$N2 = -0.48 \pm 0.03$, and therefore 12 + log(O/H) = $8.63 \pm 0.18$ (this uncertainty is dominated by the 0.18 dex systematic uncertainty
in the calibration given in Equation \ref{metallicity.eqn}), consistent with near-solar enrichment
(12 + log(O/H)$_{\odot} = 8.66$; Asplund et al. 2004).  This is consistent with the results of Shapley et al. (2004), who find
12 + log(O/H) = $8.60 \pm 0.18$ based on long-slit spectroscopy.
As noted previously by Shapley et al. (2004) however, the FWHM of the [N\,{\sc ii}] emission line is roughly 30\% larger than that of H$\alpha$, which
can be a signature of shock ionization (Lehnert \& Heckman 1996b).
Since Q1623-BX453 shows no evidence of AGN activity in either rest-frame UV spectra
(\S 4.3) or long-wavelength deviations from simple stellar population models (\S 5.1), and the measured value of $N2$ is 
consistent with that observed for normal star-forming galaxies (Shapley et al. 2004; Erb et al. 2006c) we see no concrete evidence
that this high ratio is due to shock heating or physical processes other than simple enrichment.
Erb et al. (2006c) have considered this question in greater detail using a [O\,{\sc iii}]/H$\beta$ vs. [N\,{\sc ii}]/H$\alpha$ diagnostic diagram and find that
while the [O\,{\sc iii}]/H$\beta$ ratio is greater for a given [N\,{\sc ii}]/H$\alpha$ ratio in Q1623-BX453 than in local galaxies, this offset may be explained
by a harder ionizing spectrum or a higher electron density (see also discussion by Shapley et al. 2005b) and does not provide concrete evidence for a
central AGN.

\begin{figure*}
\plotone{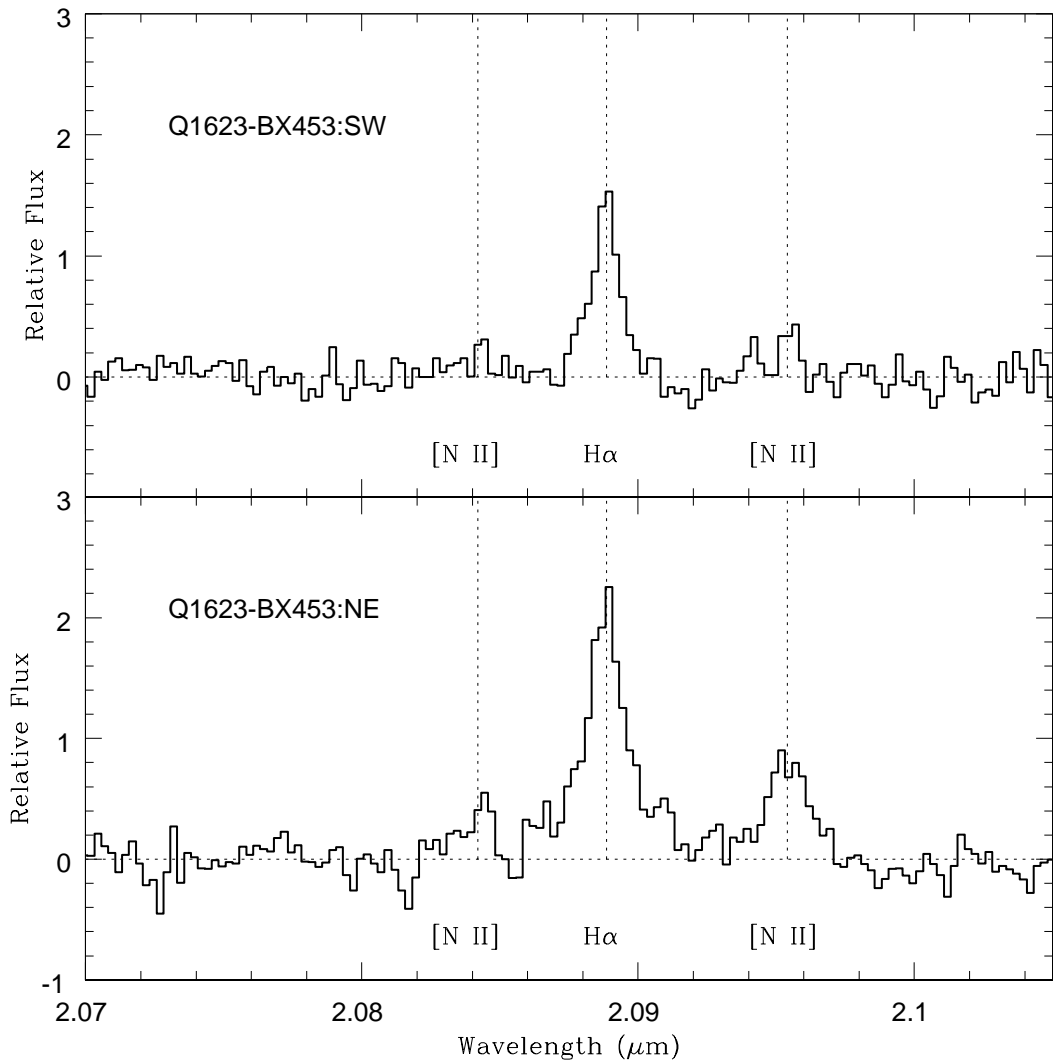}
\caption{
OSIRIS spectra of the northeastern (NE) and southwestern (SW) regions of Q1623-BX453.
Vertical dotted lines denote the wavelengths of
[N {\sc ii}] $\lambda 6550$, H$\alpha$, and [N {\sc ii}] $\lambda 6585$ emission.}
\label{twospec.fig}
\end{figure*}

Although the spectra are too faint to determine the precise [N\,{\sc ii}] morphology, we note that [N\,{\sc ii}] appears
to be coincident with the H$\alpha$ peak and varies slightly in relative strength across the face of the galaxy.
Figure \ref{twospec.fig} plots individual spectra of the northeastern and southwestern quadrants and illustrates the stronger [N\,{\sc ii}]
lines present in the northeastern region.  Taking careful measurements of the respective spectra we find (Table \ref{BX453regional.table})
that the flux ratio of the bright core is $N2 = -0.46 \pm 0.03$ (consistent with the overall spectrum;
as expected since it dominates the total flux of the source) while the flux ratio of the fainter region is
$N2 = -0.67 \pm 0.09$.  This $\sim 2.3\sigma$ difference ($\sim 0.1$ dex in log(O/H))offers mild
evidence to suggest that the distribution of heavy elements in Q1623-BX453 is not perfectly well-mixed, suggesting either 
that heavy elements from the latest burst of star formation
(produced primarily in the northeastern region) have
not yet had time to be fully distributed throughout the galaxy, or that the two regions may have had distinct star formation
histories (as might be expected for instance in a merger-type scenario).

Since Q0449-BX93 is considerably fainter than Q1623-BX453 it is not
possible to detect [N\,{\sc ii}] emission confidently.  We place a $3\sigma$ detection limit on $f_{\textrm{[N\,{\sc ii}]}}$ (Table \ref{globals.table})
and conclude that an upper bound on the metallicity is 12 + log(O/H) $\lesssim 8.37$ (i.e. less than 1/2 solar).
Given the monotonic relation between stellar mass and metallicity found by Erb et al. (2006c) for similar galaxies at $z \sim 2$,
this metallicity limit implies that $M_{\star} \lesssim 10^{10} M_{\odot}$.
Unfortunately, it is not possible to constrain the metallicity of DSF2237a-C2 as we have only observed the wavelength
interval surrounding [O\,{\sc iii}] $\lambda$5007\AA.

\begin{deluxetable}{c|cccc}
\tablecolumns{5}
\tablewidth{0pc}
\tabletypesize{\scriptsize}
\tablecaption{Q1623-BX453 Regional Emission Line Characteristics\tablenotemark{a}.}
\tablehead{
\colhead{Region} & \colhead{$f_{\rm H\alpha}$} & \colhead {$f_{\rm [N II] \lambda 6549}$} & \colhead {$f_{\rm [N II] \lambda 6585}$} & \colhead {$N2$\tablenotemark{b}}}
\startdata
NE & $14.8 \pm 0.4$ & $1.9 \pm 0.4$ & $5.1 \pm 0.4$ & $-0.46 \pm 0.03$\\
SW & $6.6 \pm 0.4$ & $< 1.0$ & $1.4 \pm 0.4$ & $-0.67 \pm 0.09$\\
\enddata
\label{BX453regional.table}
\tablenotetext{a}{Emission line flux in units of $10^{-17}$ erg s$^{-1}$ cm$^{-2}$.  Limits are 3$\sigma$ limits.}
\tablenotetext{b}{$N2 =$ log([N\,{\sc ii}]/H$\alpha$).}
\end{deluxetable}

\subsection{Outflow Properties: UV Spectroscopy}

\begin{figure*}
\plotone{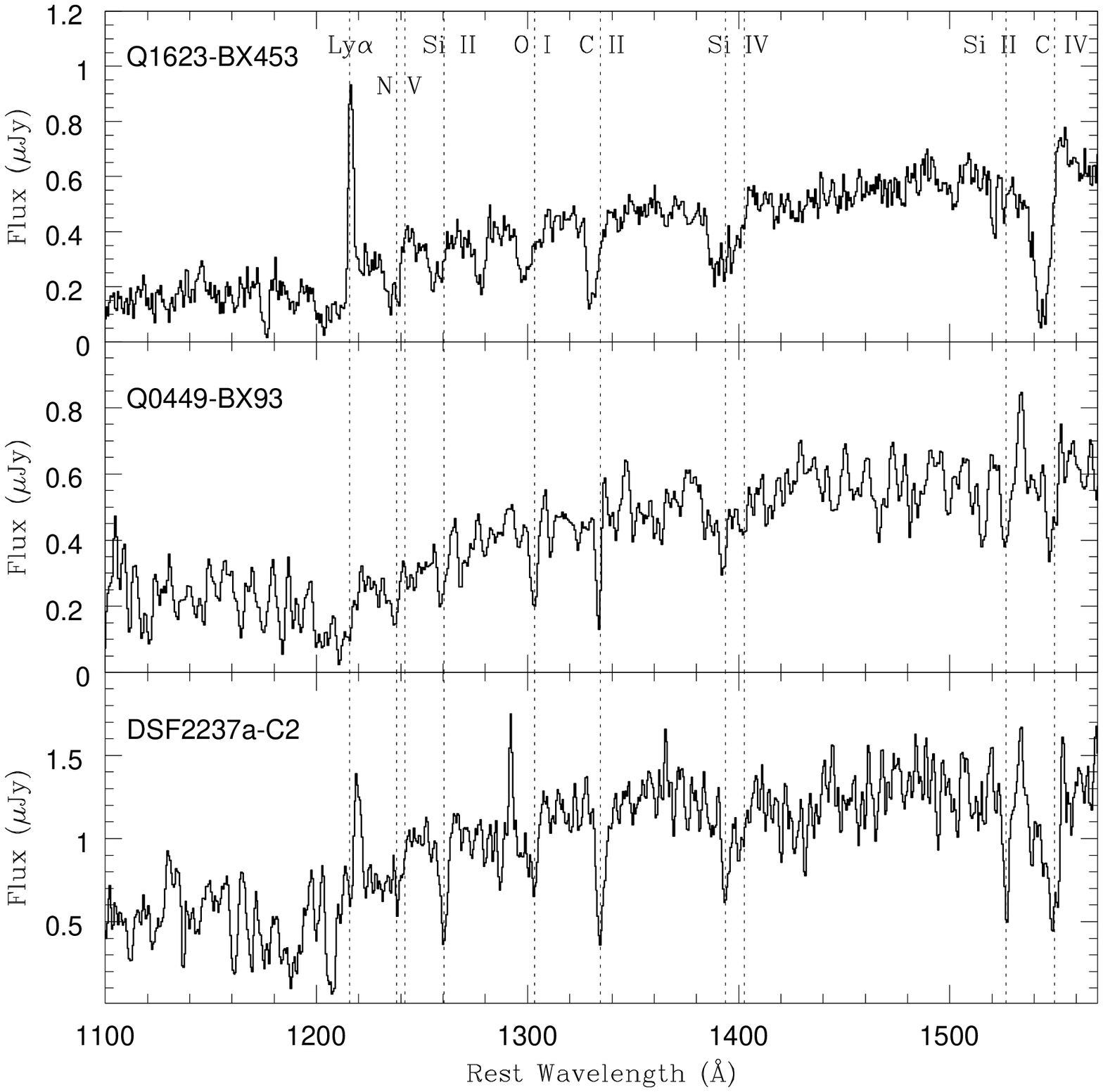}
\caption{Flux calibrated UV spectra of the three target galaxies.  Spectra have been shifted to the systemic rest frame using the nebular line redshifts
given in Table \ref{globals.table}.  The identifications and wavelengths of the strongest spectral features are indicated by vertical dotted lines.  Spectra
of Q0449-BX93 and DSF2237a-C2 have been smoothed with a 3-pixel boxcar filter.}
\label{UVspec.fig}
\end{figure*}

In Figure \ref{UVspec.fig} we plot the Keck-LRIS UV spectra of the three target galaxies
shifted into the rest-frame as defined by the OSIRIS nebular redshifts. 
The spectrum of DSF2237a-C2 was obtained as part of a survey of $z \sim3$ LBGs (Steidel et al 2003; see also
Giallongo et al 2002),
while Q1623-BX453 and Q0449-BX93 were obtained more recently using LRIS-B, described by
Steidel et al (2004).  Q1623-BX453 was included on a slit mask receiving a total of 5.5 hours integration
time, whereas the other spectra were obtained using shorter, survey-mode integration times of 1.5-2 hours. 

The spectra of all three galaxies are broadly similar to high-quality composite spectra of the $z \sim 2 - 3$ star-forming galaxy sample
(e.g. Shapley et al. 2003); these spectra exhibit strong interstellar absorption features
(e.g. Si {\sc ii}, O {\sc i}, C {\sc ii}, Si {\sc iv}, C {\sc iv}) and Lyman $\alpha$ that may appear in either
emission or absorption. It is typical of star forming galaxies at these redshifts for the interstellar
lines to be blue-shifted, and Ly$\alpha$ emission red-shifted (when present), by several hundred km s$^{-1}$ with respect to the systemic
redshift; these features are generally interpreted as scattered emission from the far-side and absorption in the near-side
of an expanding shell of outflowing gas (see, e.g., Steidel et al. 2007).
The mean values for a sample of $\simeq 100$ galaxies in the redshift range $1.9 \le z \le 2.6$ are
$\Delta v_{\rm abs}=-164$ \kms and $\Delta v_{\rm em} =+445$ \kms for the centroid of the interstellar lines and Ly$\alpha$ emission
relative to H$\alpha$ (which tends to trace gas at the systemic redshift), respectively (Steidel et al. 2007). 

The spectra of Q1623-BX453 and of Q0449-BX93 are typical with regard to the {\it sense} of the velocity offsets observed, as
summarized in Table 3. The centroids of the interstellar lines of Q0449-BX93 are blue-shifted by $\simeq -270$ \kms\ with respect
to the H$\alpha$ redshift, placing it within the top $\sim 10-15$\% in terms of this quantity, and making it very
similar to the well-studied spectrum of cB58 (Pettini et al 2000, 2002). We note in passing that relatively strong Ly$\alpha$, C\,{\sc iv}, and Si\,{\sc iv} 
absorption lines at $z_{\rm abs} = 2.0074$ are detected in the Keck/HIRES spectrum of the $z_{\rm em} = 2.677$
QSO Q0449$-$1645.  The difference from the systemic redshift of Q0449-BX93 is only 70\,km~s$^{-1}$, even though the projected separation between
the QSO and galaxy sightlines is 44'', or 373 (proper) kpc.

The far-UV spectrum of Q1623-BX453 is more remarkable,
in that the blue-shifted interstellar lines have a centroid offset of $\Delta v_{\rm abs}\simeq -900$ \kms, and a 
FWHM of $\simeq 1200$ \kms, both values being by far the
largest observed in the sample of $\simeq 100$ mentioned above.  The Ly$\alpha$ emission redshift exhibits a relatively
modest shift of $\Delta v_{\rm em}=+170$ \kms. The spectrum of Q1623-BX453 also has unusually strong high-ionization 
interstellar absorption features, including NV $\lambda\lambda 1238$,1242, and many of the absorption features 
have complex velocity structure, with two dominant components at $-600$ and $-1100$ \kms. These unusual spectral
features might lead one to suspect an AGN is involved, but there are no significant emission lines 
aside from Ly$\alpha$ in the high quality spectrum, and (as discussed further below) there is no evidence from
the full spectral energy distribution to suggest the presence of anything but stellar energy. 
In spite of the rather violent kinematics of the outflowing interstellar gas, the H$\alpha$ emission
line kinematics show no hint of being affected, presumably because any H$\alpha$ emission associated with the outflowing
material is of much lower surface brightness than what we have detected with OSIRIS.  

The spectrum of DSF2237a-C2 has typically strong interstellar absorption lines, but in this case close examination reveals that they are actually
{\it redshifted} slightly with respect to the nebular redshift, with $\Delta v_{\rm abs}\simeq +100$ \kms, while the 
Ly$\alpha$ emission line is redshifted by an unusually large $\simeq 1100$ \kms. It turns out that such kinematics, where
the interstellar line centroids are consistent with zero velocity with respect to the nebular systemic velocity,  
are observed in $\sim 15$\% of $z \sim 2-3$ galaxies (Steidel et al 2007), and it is quite typical for such objects
to have unusually large Ly$\alpha$ emission offsets.  Of possible significance is that interstellar absorption
line centroid velocities consistent with zero occur relatively frequently (5 out of 7) for galaxies in the Erb et al (2006c) sample having
measured velocity shear from NIRSPEC H$\alpha$ spectra. As discussed further by Steidel et al (2007), 
these observations suggest a connection between the presence of velocity shear (as in the case of DSF2237a-C2; see \S 6.3) and increased absorption close to
zero systemic velocity, though at present it is not possible to distinguish between orientation effects (e.g., outflows
collimated along the rotation axis) and real differences in the dynamical structure of the galaxies. In any case, OSIRIS is much more
sensitive to the measurement of velocity shear in the galaxies, and any correlation between the kinematics of
gas associated with outflows and the kinematics of the H II regions should become clearer as the sample size increases.


\section{Stellar populations, star formation rates and gas masses}

\subsection{Stellar Masses and Stellar Populations from SED Modeling}
\label{sec:seds}
As part of an ongoing survey program we have obtained broadband
photometry (Table \ref{phot.table}) in each of our target
fields.  The wavelength coverage of these data varies from field to
field: all fields (Q1623, Q0449 and DSF2237a) have ground-based optical
$U_n G {\cal R}$ photometry which serves as the basis for the
optically-selected galaxy catalog (Steidel et al. 2003, 2004); Q1623
and DSF2237a have deep near-IR $J$ and $K_s$ photometry, described by
Erb et al. (2006b) and Shapley et al. (2001) respectively; and Q1623 has
also been imaged in the mid-IR by the \textit{Spitzer Space
  Telescope}, with both IRAC at 3--8 $\micron$ and MIPS at 24
$\micron$.

\begin{deluxetable*}{lccccccccccccc}
\tablecolumns{10}
\tablewidth{0pc}
\tabletypesize{\scriptsize}
\tablecaption{Photometric Data and SED Model Parameters}
\tablehead{
\colhead{Galaxy} &\colhead{$U_n$\tablenotemark{a}} &\colhead{$G$\tablenotemark{a}} &\colhead{${\cal R}$\tablenotemark{a}} &\colhead{$J$\tablenotemark{b}} &\colhead{$K_s$\tablenotemark{b}} & \colhead{$m_{3.6\mu}$\tablenotemark{c}} & \colhead{$m_{4.5\mu}$\tablenotemark{c}} & \colhead{$m_{5.8\mu}$\tablenotemark{c}} & \colhead{$m_{8.0\mu}$\tablenotemark{c}}
& \colhead{$f_{24\mu}$\tablenotemark{d}} &\colhead{$M_{\ast}$\tablenotemark{e}} &\colhead{Age\tablenotemark{f}} &\colhead{$E(B-V)$\tablenotemark{g}}}
\startdata
Q1623-BX453 & 24.85 & 23.86 & 23.38 & 21.41 & 19.76 [20.00] & 21.38 & 21.37 & 21.33 & 21.92 & 195 & 3.1 & 404 & 0.245\\
Q0449-BX93 & 23.62 & 23.17 & 22.99 & ... & ... & ... & ... & ... & ... & ... & $\lesssim 1$\tablenotemark{h} & ...& 0.125\\
DSF2237a-C2 & ... &  24.68 & 23.55 & 22.08 & 20.53 [20.71] & ... & ... & ... & ... & ... & 2.9 & 255 & 0.175\\
\enddata
\label{phot.table}
\tablenotetext{a}{$U_n,G,{\cal R}$ magnitudes are AB.}
\tablenotetext{b}{$J,K_s$ magnitudes are Vega. Values in brackets for
  $K_s$ magnitudes represent continuum magnitudes corrected for line
  emission.}  
\tablenotetext{c}{{\it Spitzer}-IRAC magnitudes (AB standard).}  
\tablenotetext{d}{{\it Spitzer}-MIPS 24$\mu$ flux ($\mu$Jy).}  
\tablenotetext{e}{Stellar mass ($10^{10} M_{\odot}$)
  from the best-fitting constant star formation model, using the Chabrier
  (2003) IMF.  Typical uncertainty $<\sigma_{M_{\ast}}/M_{\ast}>
  = 0.4$.}  
\tablenotetext{f}{Stellar population age
  (Myr) from SED fitting, typical uncertainty $<\sigma_{\rm Age}/{\rm
    Age}> = 0.5$.}  
\tablenotetext{g}{$E(B-V)$ from the best-fit SED
  for Q1623-BX453 and DSF2237a-C2; $E(B-V)$ is estimated for
  Q0449-BX93 from the $G-{\cal R}$ color, assuming an age of 570 Myr.}
\tablenotetext{h}{Estimated from the metallicity limit (\S 4.2) using the mass-metallicity
relation of Erb et al. (2006c).}
\end{deluxetable*}

Wavelength coverage from the rest-frame UV to near-IR (in Q1623) or UV
to optical (DSF2237a) enables us to model the spectral energy
distributions of Q1623-BX453 and DSF2237a-C2 and obtain estimates of
these galaxies' stellar masses, ages, reddening and star formation
rates (the Q0449 data are restricted to the rest-frame UV and are
insufficient for such modeling).  Previous modeling of the SEDs of
these two galaxies has been presented by Erb et al. (2006b) and
Shapley et al. (2001) respectively; we redo the modeling here to add
the Spitzer IRAC data in the case of Q1623-BX453 and update the method for
consistency in the case of DSF2237a-C2.  In both cases the $K_s$ magnitudes
have been corrected for line emission according to the values
determined in \S 4.1 (the corrections are $\sim0.2$ mag in both cases).

\begin{deluxetable*}{lcccccccccc}
\tablecolumns{10}
\tablewidth{0pc}
\tabletypesize{\scriptsize}
\tablecaption{Star Formation Rates and Gas Masses}
\tablehead{
\colhead{Galaxy} & \colhead{SFR$_{\Ha}$\tablenotemark{a}} & \colhead{SFR$_{\Ha}$\tablenotemark{a}} &
\colhead{SFR$_{\rm UV}$\tablenotemark{b}} & \colhead{SFR$_{\rm UV}$\tablenotemark{b}} &
\colhead{SFR$_{\rm IR}$\tablenotemark{c}} &
\colhead{SFR$_{\rm SED}$\tablenotemark{d}} & \colhead{$\Sigma_{\rm SFR}$\tablenotemark{e}} &
\colhead{$\Sigma_{\rm SFR}$\tablenotemark{e}} &
\colhead{M$_{\rm gas}$\tablenotemark{f}} & \colhead{$\mu$\tablenotemark{g}}\\ \colhead{} & \colhead{Uncorrected} & \colhead{Corrected} &
\colhead{Uncorrected} & \colhead{Corrected} & \colhead{} & \colhead{} & \colhead{Mean} & \colhead{Peak}}
\startdata
Q1623-BX453 & 36 & 77 & 8 & 81 & 96 & 77 & 8 & 20 & 2.4 & 0.44\\
Q0449-BX93 & 12 & 18 & 12 & 39 & ... & ... & 4 & 7 & 0.7 & $\gtrsim 0.41$\\
DSF2237a-C2 & 23 & 49 & 18 & 95 & ... & 112 & 15 & 25 & 1.3 & 0.32\\
\enddata
\label{gas.table}
\tablenotetext{a}{Star formation rate ($M_{\odot}$ yr$^{-1}$)
  estimated from H$\alpha$ flux. Typical uncertainty $<\sigma_{\rm SFR}/{\rm SFR}>= 0.3$.}
\tablenotetext{b}{Star formation rate ($M_{\odot}$ yr$^{-1}$)
  estimated from rest-frame UV continuum flux.}
\tablenotetext{c}{Star formation rate ($M_{\odot}$ yr$^{-1}$)
  estimated from rest-frame 8$\mu$ flux.}
\tablenotetext{d}{Star formation rate ($M_{\odot}$ yr$^{-1}$)
  derived from SED modeling. Typical uncertainty $<\sigma_{\rm SFR}/{\rm SFR}>= 0.6$.} 
\tablenotetext{e}{Star formation rate surface density ($M_{\odot}$ yr$^{-1}$ kpc$^{-2}$).}
\tablenotetext{f}{Molecular gas mass ($10^{10} M_{\odot}$).  Typical uncertainty $<\sigma_{M}/M>= 0.2$.}
\tablenotetext{g}{Gas fraction $\mu = M_{\rm gas}/(M_{\rm gas} + M_{\ast})$. Typical uncertainty $<\sigma_{\mu}/\mu>= 0.35$.}
\end{deluxetable*}

\begin{figure*}
\plottwo{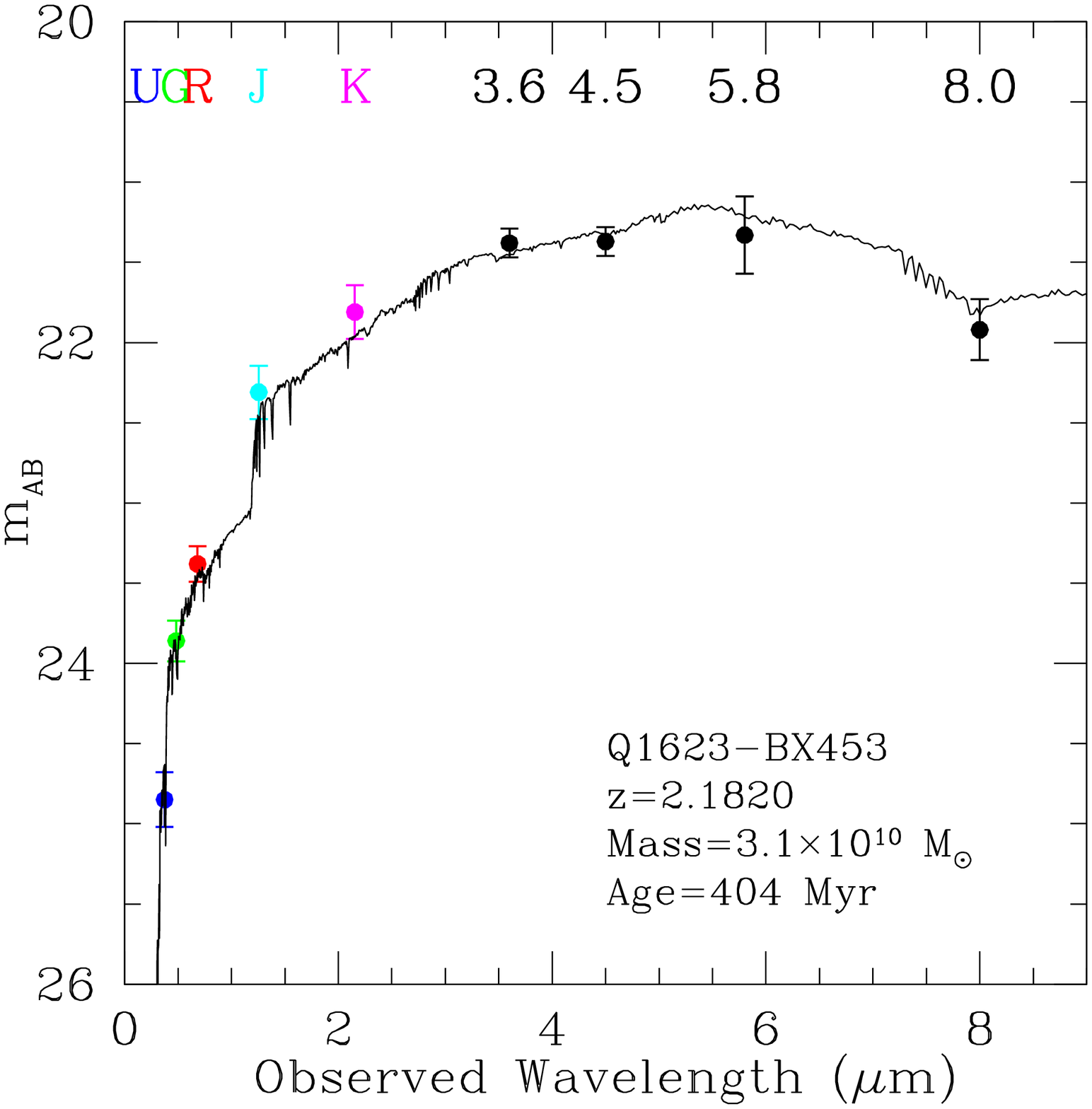}{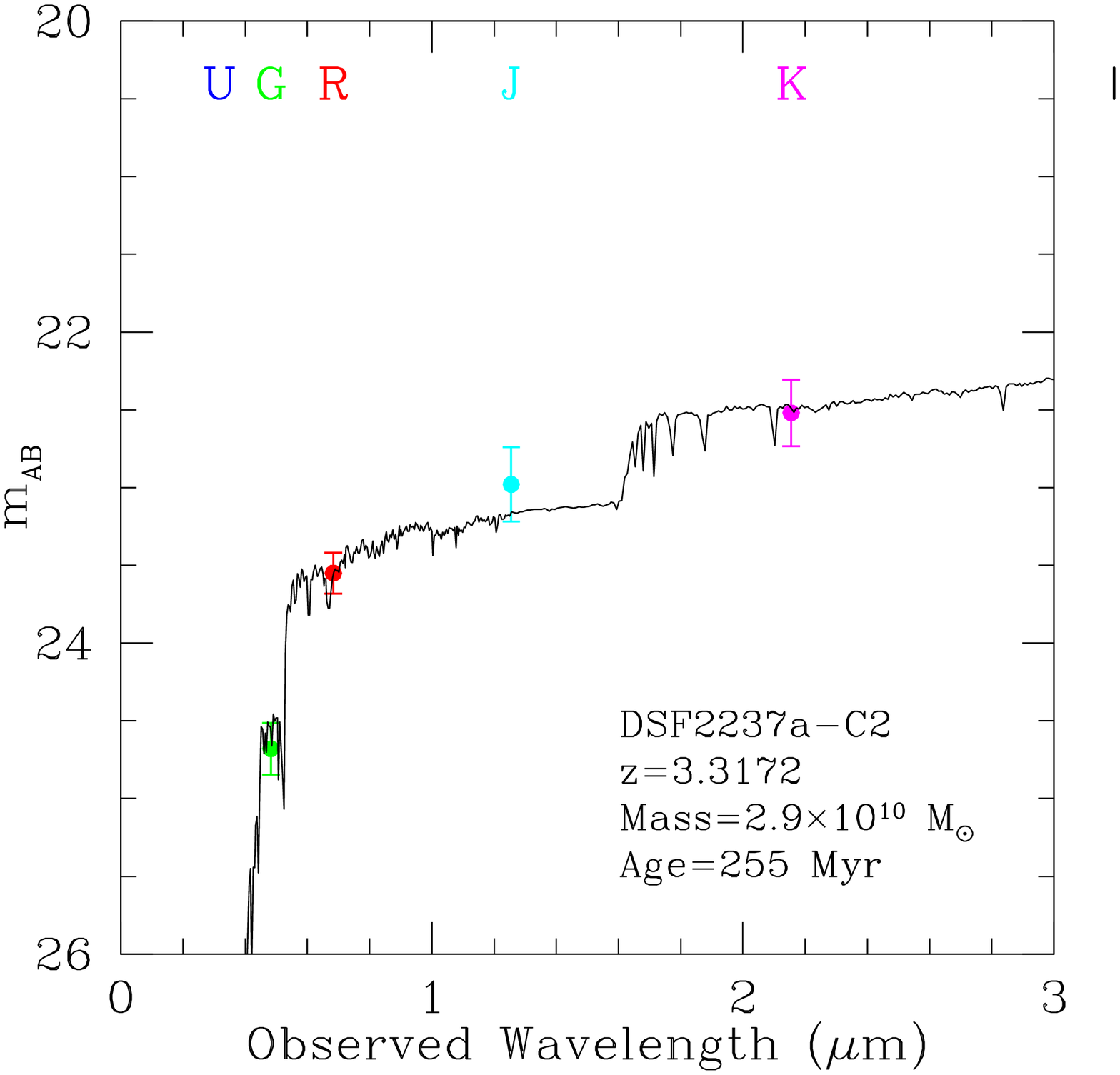}
\caption{The best-fit constant star formation (CSF) model (solid black line) is overplotted against the observed spectral energy distribution for Q1623-BX453
and DSF2237a-C2.  Colored points represent ground-based optical and near-IR photometry, black points are based on \textit{Spitzer}-IRAC observations.
The $K$-band photometry for both galaxies is shown corrected for nebular line emission (see discussion in \S 5.1).
Values given for stellar mass and population age represent the values derived from the best-fit CSF model; typical uncertainties are given
in Table \ref{phot.table}.}
\label{SEDplots.fig}
\end{figure*}

The modeling procedure has been described in detail by Shapley et
al. (2005a).  Briefly, we use Bruzual \& Charlot (2003) models with
a constant star formation rate, solar metallicity, and a Chabrier (2003) IMF.
The resulting stellar masses for Q1623-BX453 and DSF2237a-C2 are
$M_\star = 3.1\times10^{10}$ \msun\ and $M_\star = 2.9\times10^{10}$
\msun\ respectively.  The remaining stellar population parameters are
given in Table~\ref{phot.table}, and the best-fit spectral energy
distributions are shown in Figure~\ref{SEDplots.fig}.

\subsection{Star Formation Rates}

A direct estimate of the instantaneous star formation rate may be derived
from the observed H$\alpha$ luminosity since it traces the
ionizing radiation produced by young, massive stars.  We adopt the
calibration derived by Kennicutt et al.  (1994; summarized in
Kennicutt 1998) whereby
\begin{equation}
\textrm{SFR}\;(\msunyr) = \frac{L(\Ha)}{1.26\times 10^{41} \; \textrm {erg s}^{-1}} \; \times 0.56 .
\end{equation}
where we have introduced the factor of 0.56 to convert to the Chabrier (2003) IMF.

This enables direct calculation of the star formation rate from our
observed \Ha\ fluxes for Q1623-BX453 and Q0449-BX93; we find
SFR$_{\Ha}=36$ \msunyr\ for Q1623-BX453 and SFR$_{\Ha}=12$
\msunyr\ for Q0449-BX93.  Although H$\alpha$ is redshifted beyond the
$K$-band for $z \gtrsim 2.6$, it is nonetheless possible to estimate a
star formation rate for DSF2237a-C2 based on the observed
[\ion{O}{3}] flux.
Using long-slit spectroscopy, Pettini et al. (2001) determined the
H$\beta$ flux to be $3.5 \pm 0.4 \times 10^{-17}$ erg s$^{-1}$
cm$^{-2}$, and the ratio of the [\ion{O}{3}] and
\Hb\ fluxes to be $F_{\Hb}/F_{\rm [O\,III]}=0.19$ (their
[\ion{O}{3}]$\lambda 5007$ flux of $18 \times 10^{-17}$ erg s$^{-1}$
cm$^{-2}$ is in rough agreement with the OSIRIS value of $10.1
\pm 0.5 \times 10^{-17}$ erg s$^{-1}$ cm$^{-2}$, given the significant
uncertainties in the calibration of long-slit spectra).  
While \Hb \, falls outside the narrow wavelength range which we have observed
with OSIRIS, we can use the $F_{\Hb}/F_{\rm [O\,III]}$ ratio observed by Pettini
et al.\ (2001) to infer an \Hb\ flux of $2.0\times 10^{-17}$ erg
s$^{-1}$ cm$^{-2}$ from our observed [\ion{O}{3}] value.
We then assume a standard recombination ratio of $F_{\Ha}/F_{\Hb} =
2.75$ (Osterbrock 1989) to find $F_{\Ha} = 5.4 \times 10^{-17}$ erg
s$^{-1}$ cm$^{-2}$ and SFR$_{\Ha}=23$ \msunyr.  These SFRs are
uncorrected for dust extinction (which we discuss below) and have
typical uncertainties $\sim$ 30\% (dominated by the uncertainty in the flux
calibration described in \S 4.1).

We use the values of $E(B-V)$ determined by the SED modeling discussed
in \S\ref{sec:seds} to correct the observed \Ha\ flux of Q1623-BX453
and the inferred \Hb\ flux of DSF2237a-C2 for extinction
using a modified Calzetti et al. (2000) starburst
attenuation law as described by Erb et al. (2006b).
For Q1623-BX453 we
find $E(B-V)=0.245$ for a corrected SFR of 77 \msunyr, in agreement
with the value of 77 \msunyr\ we obtain from the SED modeling.  For
DSF2237a-C2 we use the best fit $E(B-V)=0.175$ to correct the \Hb\ flux
inferred above, and then assume $F_{\Ha}/F_{\Hb} = 2.75$ to
find $F_{\Ha}=11.3 \times 10^{-17}$ erg s$^{-1}$ cm$^{-2}$ and
SFR$_{\Ha}=49$ \msunyr.  This is somewhat smaller than the value of
112 \msunyr\ we find from the SED modeling, but not inconsistent with
the factor of $\sim2$ RMS found by Erb et al. (2006b) for SFRs
determined from SED modeling and \Ha\ emission, especially given the
additional extrapolations needed to calculate the nebular line SFR.\footnote{Curiously, the original
H$\beta$ flux measured by Pettini et al. (2001) gives a better match to SED star formation rate
than the extrapolated flux measured using OSIRIS, highlighting the possibility that OSIRIS may 
sometimes `miss' low surface brightness flux.}

Because we lack the IR data to model the SED of Q0449-BX93, we
estimate the extinction from the $G-{\cal R}$ color alone, assuming a
constant star formation age of 570 Myr, the median age\footnote{The values of $E(B-V)$ allowed by the $G-{\cal R}$ color
are relatively insensitive to the age of the galaxy; for reasonable
ages ranging from 50 Myr (approximately the dynamical timescale) to
3.2 Gyr (the age of the universe at the redshift of the galaxy),
$E(B-V)$ varies from 0.164 to 0.113.} of the
\ztwo\ sample as found by Erb et al. (2006a).
Assuming this median age we find $E(B-V)=0.125$ and
a corrected SFR of 18 \msunyr.  The uncertainty in age has a
negligible effect on the corrected SFR; for an age of 50 Myr, we find
SFR$_{\Ha}=20$ \msunyr, while the maximum age of 3.2 Gyr leads to
SFR$_{\Ha}=17$ \msunyr.  The corrected and uncorrected SFRs for the
three galaxies are summarized in Table~\ref{gas.table}, and for
reference we note that the mean extinction-corrected SFR of the
\ztwo\ sample, from both \Ha\ and the UV continuum, is $\sim30$
\msunyr\ (Erb et al. 2006a).

By way of comparison, we also include in Table~\ref{gas.table} estimates of the star
formation rate obtained using UV and mid-IR photometry.  In the former case,
we use the rest-frame monochromatic luminosity at 1500\AA ($L_{\nu}$; based on linear
interpolation of the broadband photometry given in Table \ref{phot.table}) and the Kennicutt (1998b) conversion between $L_{\nu}$ and SFR:
\begin{equation}
\textrm{SFR}\;(\msunyr) = \frac{L_{\nu}}{7.14 \; \times \; 10^{27} \; \textrm{erg s}^{-1} \; \textrm{Hz}^{-1}} \; \times 0.56
\end{equation}
where the factor of 0.56 has again been introduced to convert to the Chabrier (2003) IMF.
After correcting for extinction using the Calzetti et al. (2000) starburst attenuation law, we find 
SFR$_{\rm UV}$ = 81, 39, 95 \msunyr \, for Q1623-BX453, Q0449-BX93, and DSF2237a-C2 respectively.  
These values are in good agreement with SFR$_{\rm H\alpha}$ for Q1623-BX453; the factor of $\sim 2$ difference between UV and H$\alpha$ estimates
for DSF2237a-C2 and Q0449-BX93 may be due (in the former case) to the uncertainties introduced in our conversion from [O {\sc iii}] flux to H$\alpha$
and suggests (in the latter case) that Q0449-BX93  may be a particularly young, blue, dust-free galaxy for which
the conversion from $(G-{\cal R})$ color to $E(B-V)$ is imperfect.
Using {\it Spitzer}-MIPS 24$\mu$ photometry, it is also possible to estimate the SFR for Q1623-BX453 using the rest-frame 8$\mu$ flux
given in Table \ref{phot.table}.  Adopting the calibrations discussed in Reddy et al. (2006), we estimate SFR$_{\rm IR} =$ 96 $M_{\odot}$ yr$^{-1}$
under the Chabrier (2003) IMF, in good agreement with the UV, SED, and H$\alpha$ estimates.

\subsection{Gas Masses and Gas Fractions}

We can also use the \Ha\ luminosities in combination with the galaxy
sizes measured in \S 3 to estimate the galaxies' gas masses and gas
fractions. Such indirect techniques are required because emission from
molecular gas is very difficult to detect in all but the most luminous
high redshift galaxies with current technology.  We follow the
procedure described by Erb et al. (2006a), benefitting here from the
higher spatial resolution of the OSIRIS data, which provides
better constraints on the galaxies' sizes.  
As described by Erb et al. (2006a), we use the empirical
global Schmidt law (Kennicutt 1998), which relates the surface density
of star formation to the surface density of molecular gas (in the
starburst regime in which our galaxies fall; more quiescent
star-forming galaxies at low redshift also have significant amounts of
atomic gas).

We determine the surface density of H$\alpha$ luminosity ($\Sigma_{\Ha}$) by dividing the extinction-corrected
observed (or inferred, in the case of DSF2237a-C2) \Ha\ luminosity by the
projected linear area $I$ of the galaxy as determined in \S 3.  The
gas surface density $\Sigma_{\rm gas}$ is then given by the global Schmidt law, and
the total gas mass by $M_{\rm gas}=\Sigma_{\rm gas} I$.  Gas masses
inferred in this way are subject to significant uncertainties beyond
those associated with the measurements of fluxes and sizes and the
$\sim0.3$ dex uncertainty introduced by the scatter in the Schmidt law
itself.  It is not yet known whether the Schmidt law in this form
holds at high redshift, although the one galaxy for which it has been
tested, the gravationally lensed Lyman break galaxy MS1512-cB58,
appears to be consistent with the local form (Baker et al.\ 2004).  It
is also likely that the Schmidt law, which applies to only the cold
gas associated with star formation, does not account for all of the
gas; in a typical local disk galaxy, $\sim40$\% of the total gas mass
is not included by the Schmidt law (Martin \& Kennicutt 2001).

The measured areas of Q1623-BX453, Q0449-BX93, and DSF2237a-C2 are $I =$ 9.5 $\pm$ 0.3,
4.7 $\pm$ 0.6, and 3.3 $\pm$ 0.3 kpc$^2$ respectively.  The size of Q1623-BX453 derived
here ($r = 1.74 \pm 0.03$ kpc) is considerably smaller than the value of $r = 4.4 \pm 0.4$
found by Erb et al. (2006b) from long-slit NIRSPEC observations, reflecting in part the slightly different definitions
of $r$ and the difficulty of estimating sizes from
seeing-limited long-slit data.  However, it is also likely that the luminous
area of a given galaxy is genuinely greater in the Erb et al. (2006b) data since NIRSPEC is more sensitive
than OSIRIS, and may thus detect lower surface-brightness features at larger radii.  Galaxy size
$r$ and $I$ as defined here are only meaningful with regard to the 
detection threshhold of a particular study.

Combining these size estimates
with the extinction-corrected \Ha\ luminosities, we find
gas masses of 2.4 $\times10^{10}$, 6.9 $\times10^{9}$, and 1.3 $\times10^{10}$
\msun 
(each with $\sim$ 20\% statistical uncertainty, but note discussion of systematic uncertainties above) for Q1623-BX453, Q0449-BX93, and DSF2237a-C2
respectively.  The stellar mass of Q1623-BX453 is
3.1$\times10^{10}$ \msun, for a gas fraction $\mu=0.44$; for
DSF2237a-C2 we find $M_{\star}=1.6\times10^{10}$ \msun and $\mu=0.32$.
We are unable to determine a gas fraction for Q0449-BX93 because of
the lack of near-IR imaging from which to derive a stellar mass, although the mass limit estimated from the
mass-metallicity relation (Erb et al. 2006c) suggests that $\mu \gtrsim 0.41$.
Combining the typical uncertainties of the stellar ($\sigma_{\rm M_{\star}}/{\rm M_{\star}} = 0.4$) 
and gas ($\sigma_{\rm M_{\rm gas}}/{\rm M_{\rm gas}} = 0.2$) masses,
we estimate the uncertainty in the gas fraction $\sigma_{\mu}/{\mu} = 0.35$.
The stellar mass and gas fraction of
Q1623-BX453 are close to the mean values ($<M_{\star}> = 3.6 \pm 0.4 \times 10^{10} M_{\odot}$, $<\mu> = 0.37 \pm 0.03$) found for a larger sample
of 114 galaxies at $z\sim2$ by Erb et al. (2006b).
While there is no similar determination of the mean gas fraction at $z \sim 3$, the stellar mass of DSF2237a-C2
is only slightly higher than
the median of $1.4\times10^{10}$ \msun  found for $z\sim3$ Lyman break galaxies by
Shapley et al. (2001; we convert their reported
median to our adopted IMF).

We also use the extinction-corrected SFRs determined above to estimate
the average and peak star formation rate surface density $\Sigma_{\rm SFR}$ (Table \ref{gas.table}).  Dividing
the SFRs by the total areas, we find an average $\Sigma_{\rm SFR}$ = 8, 4, and 15
\msunyr\ kpc$^{-2}$ for Q1623-BX453, Q0449-BX93 and DSF2237a-C2
respectively.
These values lie in the middle of the distribution of
$\Sigma_{\rm SFR}$ found for local starbursts by Kennicutt 1998 (after
accounting for the different IMF), and at the upper end of the range
for \ztwo\ galaxies found by Erb et al. (2006b).
Based on the peak flux densities observed in a given spaxel, we estimate the corresponding peak star formation rate surface densities to be
roughly twice the mean, with
$\Sigma_{\rm SFR}$ = 20, 7, and 25 \msunyr\ kpc$^{-2}$.
These values are comparable to the maximal star-formation rate densities observed in the local universe by
Lehnert \& Heckman (1996a; $\Sigma_{\rm lim}$ = 11 \msunyr\ kpc$^{-2}$ in our adopted IMF) and Meurer et al. (1997;
$\Sigma_{\rm lim}$ = 25 \msunyr\ kpc$^{-2}$ in our adopted IMF), implying that Q1623-BX453 and DSF2237a-C2 may be forming
stars at close to the maximal rate in their brightest regions.
Given that (in the local universe) a SFR density of $\Sigma_{\ast} = 0.1 M_{\odot}$ yr$^{-1}$ kpc$^{-2}$ is sufficient for the resulting supernovae 
to drive a wind into the surrounding ISM (Heckman 2002), 
it is likely that rapidly star forming galaxies similar to these are quite ``leaky'' systems, with a significant fraction
of metal-enriched gas finding its way into the IGM (Erb et al. 2006c; Adelberger et al. 2003,2005b).


\section{SPATIALLY RESOLVED KINEMATICS AND DYNAMICAL MASSES}

Using the integrated spectra described in \S 4.1 we estimate the kinematic dispersion of the ionized
gas by fitting a gaussian profile (corrected for instrumental resolution) to the observed emission.  We obtain values of $\sigma_{\rm v} =
92 \pm 8, 72 \pm 12, 101 \pm 15$ km s$^{-1}$ respectively for Q1623-BX453, Q0449-BX93, and DSF2237a-C2.

Although our velocity dispersion for Q1623-BX453
is slightly higher than that obtained previously by Erb et al. (2006b; $\sigma_{\rm v} = 61 \pm 4$ km s$^{-1}$)\footnote{This discrepancy 
appears to be due to the presence of a residual OH sky feature in the NIRSPEC H$\alpha$ spectrum; the OSIRIS velocity dispersion is roughly in agreement with
the NIRSPEC velocity dispersion of $\sigma_{\rm v} = 106 \pm 8$ km s$^{-1}$ measured using the [O\,{\sc iii}] emission line.}, the measured dispersion
of DSF2237a-C2 agrees
with the value $\sigma_{\rm v} = 100 \pm 4$ km s$^{-1}$ observed by Pettini et al. (2001).  However, as we demonstrate below
in \S 6.3 the 
velocity shear of DSF2237a-C2 contributes to the line width of the integrated spectrum.
Based on examination of
the spatially resolved velocity map of the galaxy, we conclude that the central velocity dispersion $\sigma_{\rm v} = 79$ km s$^{-1}$, rising
slightly to about $89$ km s$^{-1}$ at larger radii.

While the total amplitude of any velocity shear may be an unreliable probe of the dynamical mass of a system, the
central velocity dispersion $\sigma$ is predominantly driven by the gravitational potential of the system even in kinematically complex
systems such as local ULIRGS
(Colina et al. 2005).  
These $\sigma$ may therefore be used to estimate the dynamical mass of the galaxies using the relation
\begin{equation}
M_{\rm dyn} = \frac{C \sigma^2 r}{G}
\label{dynmass.eqn}
\end{equation}
where a relatively straight-forward derivation (e.g. Erb et al. 2006b) 
gives the constant geometric prefactors $C = 5$ appropriate for a uniform sphere and $C = 3.4$ for a thin disk
with average inclination.
Based on the kinematic data presented in \S 6.1 - 6.3, we adopt $C = 5$ for Q1623-BX453 and Q0449-BX93, and $C = 3.4$ for DSF2237a-C2.

While $\sigma$ is reasonably well-determined, $r$ is not and can affect the resulting dynamical mass 
calculated via Eqn. \ref{dynmass.eqn} significantly.
Ideally, we might adopt the virial
radius, for which the derived $M_{\rm dyn}$ should be a reasonable estimate of the total gravitational mass.
However, the only observationally-determined length scale is that of the much smaller region
from which nebular line emission is detected (see Table \ref{OSIRISmorphs.table}).
Adopting this nebular emission radius, we calculate dynamical masses of 
$M_{\rm dyn}$ = 17 $\pm$ 2, 7 $\pm$ 2, and 5 $\pm$ 1 $\times 10^{9} M_{\odot}$ for Q1623-BX453, Q0449-BX93, and DSF2237a-C2 
(using $\sigma_{\rm v} = 79$ km s$^{-1}$)
within a radius $r =$ 1.74, 1.23, and 1.03 kpc respectively.  Unsurprisingly (given that our dynamical masses probe only the central
1-2 kpc), these masses differ by $\sim$ a factor of 15 from the typical halo
mass within the virial radius calculated for similar galaxies
by Adelberger et al. (2005a) based on clustering statistics.  A previous estimate of $M_{\rm dyn}$ has also
been obtained for Q1623-BX453 by Erb et al. (2006b); their value $M_{\rm dyn} = 12 \times 10^{9} M_{\odot}$ is similar to our own, with the predominant
difference arising from the difference in observed $\sigma_{\rm v}$.

We construct kinematic maps of the three target galaxies by fitting gaussian profiles to the spectra contained within each resampled spaxel.
Generally, single-component models suffice to represent the emission line profiles (although
see discussion in \S 6.2 for Q0449-BX93).
Such maps of the relative radial velocity and velocity dispersion 
(Fig. \ref{resultsfig.fig}, center and right panels respectively) are intrinsically noisy; 
we reduce the contribution of noisy spaxels to the resulting maps
by considering only those for which the derived velocity is within 300 km s$^{-1}$ of the systemic redshift, the
FWHM is greater than or equal to the instrumental width, and the SNR of the detection is greater than six.  These criteria
are not derived from theoretical limits, but rather represent empirical constraints which we have found (based on iterations with various parameters)
to produce the cleanest kinematic maps.
These velocity maps are generally robust to systematic uncertainties (since all spaxels are processed in a similar manner) and
the largest source of uncertainty is from random errors arising from the difficulty in determining
the accurate centroid for an emission line of given S/N ratio.
Given the OSIRIS spectral resolution $R \sim 3600$, the velocity FWHM is about 80 km s$^{-1}$.
We estimate that uncertainties in the velocity maps are 
$\sim$ 10 km s$^{-1}$ for bright central regions of a galaxy, tapering to $\sim 20$ km s$^{-1}$ towards the edges.

\subsection{Q1623-BX453}

The 2d velocity map of Q1623-BX453 (Fig. \ref{resultsfig.fig}, top middle panel) is largely consistent with a flat velocity field across the entire
galaxy ($\sim$ 4 kpc along the major axis) and does not present any evidence for spatially resolved shear.
Typical deviations from zero relative velocity are
within the uncertainty of our ability to accurately measure
velocity centroids from individual spaxels (i.e. about 10-20 km s$^{-1}$).
The only suggestion of kinematic substructure is a small $\sim$ 100 mas (800 pc) feature at the northeastern edge of the galaxy
which appears to be receding at $\sim$ 50 km s$^{-1}$ relative to systemic and has a sharp discontinuity with the bulk
velocity field (i.e. it does not appear to ``blend in'' by means of a gradual velocity transition).
Given the extremely low flux of this feature (which lies outside the faintest flux contour illustrated on Fig. \ref{resultsfig.fig}) 
this may be simply a residual noise artifact.

Although the line-of-sight velocity of the ionized gas is roughly constant across the face of the galaxy, the dispersion of
this gas ($\sigma$) varies slightly with position (Fig. \ref{resultsfig.fig}; top right panel).  
In particular, $\sigma$ peaks around 100 km s$^{-1}$ in the northeastern
region approximately coincident with the flux peak, decreases to $\sim$ 70 km s$^{-1}$ in the southwestern region, and drops to below 30 km s$^{-1}$ in the 
extremely faint emission region to the west.  While there appear to be peaks of $\sigma \sim 120$ km s$^{-1}$ on the northern and eastern edges of the galaxy,
it is uncertain whether these represent real features or are simply noise in the extremely faint surface brightness edges of the galaxy.

The flat velocity field of this galaxy is particularly intriguing in light of the semi-analytic picture of galaxy formation in the early universe
(e.g. Mo, Mau, \& White 1998, Baugh 2006) in which (in brief) cold gas falls into collapsed dark halos, shock heats, radiatively cools to form a disk
supported by residual angular momentum, and proceeds to form stars within this gaseous disk.
Unless Q1623-BX453 happens to be inclined almost perfectly face-on to our line-of-sight\footnote{In order to mask a 100 km s$^{-1}$ circular velocity
within a 20 km s$^{-1}$ uncertainty a galaxy must have inclination $i \lesssim 10^{\circ}$.  This occurs in less than 2\% of cases for a randomly
distributed set of inclinations.}, the absence of a velocity gradient across the galaxy indicates that
the kinematics of the ionized gas are not dominated by ordered rotation.
Clearly, the simplest description of galaxy formation is incomplete.

One variant on the basic rotating disk
model which provides a reasonable explanation of some of the observed characteristics is that described by Noguchi et al. (1999) and 
Immeli et al. (2004; their models A/B), in which
highly-efficient cooling mechanisms lead to early fragmentation of the gas disk into self-gravitating clouds.  These clouds are sufficiently massive that 
they spiral to the center
of the galaxy via dynamical friction within a few dynamical times, largely dragging the existing stellar population along with the gas.  
Immeli et al. (2004)
find that an intense burst of star formation typically occurs in the nuclear region of the galaxy when these massive gas clouds collide,
typically peaking at around 100 $M_{\odot}$ yr$^{-1}$ (over twice that expected for ``traditional'' star formation histories in galaxies with less 
efficient gas cooling mechanisms) and leading to the rapid formation of a stellar bulge.
Such a theory may provide a natural explanation for Q1623-BX453, which we infer to have a large cold gas fraction, near-peak star formation
rate density, and a sizeable stellar population.  Alternatively, such characteristics may also be explained by the merger of two massive,
gas-rich galaxies whose cold gas reservoirs are undergoing rapid star formation in the center of the new gravitational potential.

While it might appear surprising that a system with such violent dynamics should show so little evidence for kinematic structure, such systems 
are well-known in the local universe.
Local ULIRGS may often have gas masses constituting the bulk of their dynamical mass (Solomon et al. 1997), fueling intense starbursts from the
dense molecular gas in their cores.  The common presence of multiple nuclei and tidal features (e.g. Sanders et al. 1988, Bushouse et al. 2002) suggests that 
these starbursts may commonly arise from the major merger of two gas-rich systems, and the ionized gas kinematics typically show star formation in
gas flows rapidly forming stars in the center of the gravitational potential.
The velocity fields of ULIRGS are typically dominated by strong, asymmetric gas flows (Colina et al. 2005), although the specific properties of these flows
vary considerably from case to case.  The velocity structure of Q1623-BX453 resembles in particular the kinematics of
Mrk 273 and IRAS 15250+3609 (see Fig. 1 of Colina et al. 2005); these galaxies show no strong kinematic substructure in the brightest regions of H$\alpha$
emission within $\sim 2-3$ kpc of the peak H$\alpha$ emission despite
relatively strong ($\sim 200$ km s$^{-1}$) kinematic features in regions with 
H$\alpha$ surface brightness lower by a factor $\sim$ 5.
It is an interesting speculation therefore whether deeper observations of Q1623-BX453 will reveal similar kinematic
structure in regions of lower surface brightness. 

\subsection{Q0449-BX93}

\begin{figure}
\plotone{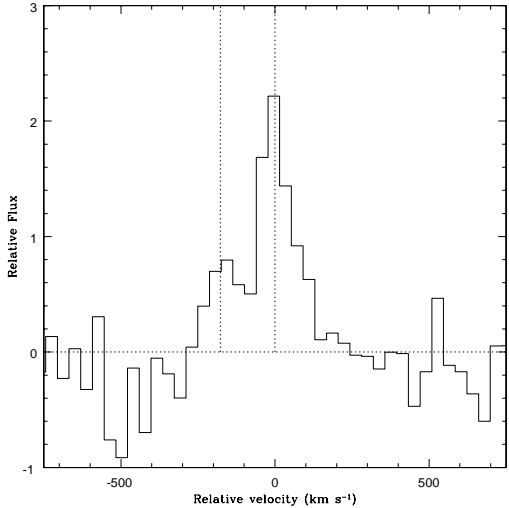}
\caption{The spectrum of Q0449-BX93 is shown (in units of velocity relative to systemic) integrated over the region in which a secondary H$\alpha$ emission
component is observed blueward of the systemic emission.  Vertical dashed lines denote the fiducial galaxy velocity (i.e. 0 km s$^{-1}$)
and the location of the secondary emission peak after deblending.  This peak occurs at a relative velocity of -177 $\pm 11$ km s$^{-1}$.}
\label{bx93_bluebump.fig}
\end{figure}

While the integrated spectrum of Q0449-BX93 (Fig. \ref{spectra.fig}) is dominated by a single emission spike defining the systemic redshift, the spectrum 
of the northwestern region
of the galaxy shows a faint secondary emission peak blueshifted by $\sim 180$ km s$^{-1}$ from the systemic redshift
(Fig. \ref{bx93_bluebump.fig}).  Given that the FWHM of the primary
emission component is $\sim 150$ km s$^{-1}$, these two components in Figure \ref{bx93_bluebump.fig} are blended, motivating the use of a 2-component
gaussian model in order to adequately characterize the emission line profile of the source.

Considering first the primary spectroscopic component,
the velocity and dispersion structure of Q0449-BX93 
(Fig. \ref{resultsfig.fig}, center and middle-right panels respectively)
are similar to those of Q1623-BX453 although
there is mild evidence for velocity shear along the northwest-southeast axis.  The peak-to-peak amplitude of this velocity differential
is only $\sim 40$ km s$^{-1}$, and therefore not highly significant with respect to our estimated velocity uncertainty of 10 - 20  km s$^{-1}$.
Even if this shear is genuine, it is not obvious that this implies a rotationally supported system.  Assuming a simplistic
model in which the circular velocity at the outer edge of the galaxy (1.23 kpc; see Table \ref{OSIRISmorphs.table}) 
is 40 km s$^{-1}$ (i.e. with an extremely generous inclination correction
to the observed 20 km s$^{-1}$), we obtain a dynamical mass of only $M_{\rm dyn} = 5 \times 10^{8} M_{\odot}$, 
far less than the gas mass alone within the same region as derived
from the H$\alpha$ emission flux.  That is, the observed shear is insufficient to provide dynamical
support for the galaxy.
As in the case of Q1623-BX453, the velocity dispersion is considerably larger ($\sigma = 72$ km s$^{-1}$) 
than the peak-to-peak velocity shear, suggesting the 
physical interpretation that Q0449-BX93 may also represent a merger-induced starburst or
the bright central emission region of a gas-rich disk which has fragmented due to instability.
While insufficient broadband data exist for Q0449-BX93 to be able to reliably fit stellar population models, 
estimates of the stellar mass using the mass-metallicity relation of Erb et al. (2006c) suggest that this galaxy
is considerably less massive ($M_{\star} \lesssim 10^{10} M_{\odot}$) than Q1623-BX453.  
Coupled with its moderate extinction-corrected H$\alpha$ star formation rate of 18 $M_{\odot}$ yr$^{-1}$
(close to the average for the $z \sim 2$ galaxy sample, Erb et al. 2006a) Q0449-BX93 may be more representative of the general
galaxy population.

\begin{figure*}
\plotone{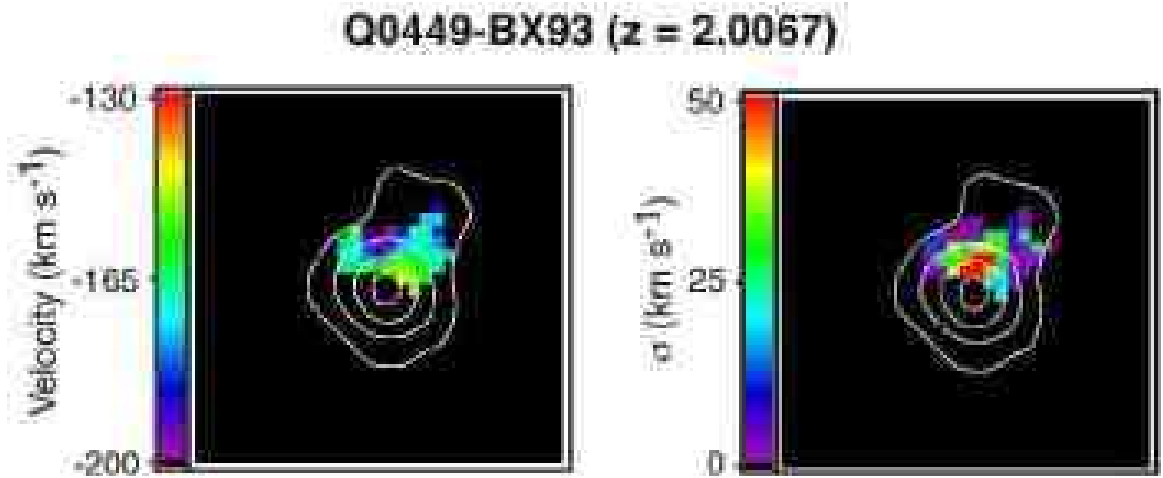}
\caption{As Figure \ref{resultsfig.fig}, shows the H$\alpha$ emission line velocity and velocity dispersion maps for the faint blueshifted component of 
Q0449-BX93.  Overlaid contours are H$\alpha$ flux density.}
\label{bx93_component2.fig}
\end{figure*}

Selecting next the gaussian fits to the secondary emission line component it is possible to create
velocity and dispersion maps (Fig. \ref{bx93_component2.fig}) of this faint, kinematically distinct sub-component (hereafter KDSC).
To the limiting surface brightness of our study, the KDSC is marginally resolved over only a small ($\sim 200 \times 150$ mas, i.e. $\sim$ 1 kpc) part
of the northwestern region of the galaxy comparable in size to the PSF ($\sim 150$ mas).
The kinematic offset of the KDSC from the systemic redshift is approximately constant across the region of emission and does not smoothly
join to the kinematics of the primary component.
While this may be simply due to the small size of the KDSC (across which we expect only $\sim 1-2$ spatially independent spectra), 
an additional interpretation (within the framework of a fragmenting disk model) may be that the KDSC 
represents one of multiple starbursting regions in the process of coalescing in the center of the galaxy for which
we happen to see a kinematic differential.
Alternatively, the KDSC may have initially begun star formation as a unit (i.e. satellite galaxy) 
distinct from the bright component of Q0449-BX93 which is only
now merging with its host galaxy and is perhaps responsible for the observed burst of star formation.
It is unfortunately beyond the scope of the current data to derive the parameters (e.g. relative mass)
of such a scenario with any pretense of reliability.

\subsection{DSF2237a-C2}

The kinematic map of ionized gas in DSF2237a-C2
is notably different from those of Q1623-BX453 and Q0449-BX93, presenting clear evidence for resolved
velocity shear along the morphological major axis at redshift $z > 3$.
As illustrated in Figure \ref{resultsfig.fig} (bottom center panel), the velocity of peak [O\,{\sc iii}] emission relative to the systemic redshift
varies smoothly from about $-70$ to $+70$ km s$^{-1}$ over a distance $\sim 2.5$ kpc.
An apparent ridge-line of high velocity (most evident in the southwest corner of the galaxy) presents some evidence
for a flattening in the velocity curve, although this is far from robust.
The velocity dispersion of individual regions within the galaxy is $\sigma \sim 79$ km s$^{-1}$; this is inflated by velocity shear
in the integrated spectrum, 
giving $\sigma = 101$ km s$^{-1}$.

\begin{figure}
\plotone{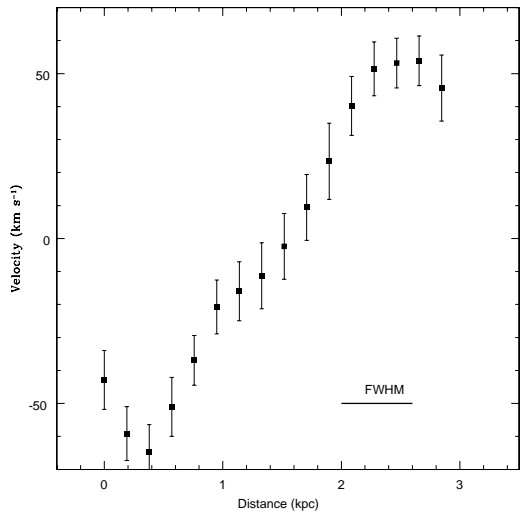}
\caption{One-dimensional spectrum of DSF2237a-C2 (solid boxes) extracted from the OSIRIS data using an idealized ``slit''.  Uncertainties are conservative estimates
based on the number of spaxels (each with 20 km s$^{-1}$ uncertainty) included in the spectrum at each position along the simulated slit.  The solid line
indicates the FWHM of the PSF.  Note that this figure underestimates the turnover velocity apparent in 2-d data because
spatial averaging across the simulated slit combines peak ridgeline velocities with lower velocities on either side of the ``v''-shaped pattern 
(see Fig. \ref{c2_diskmodel.fig}).}
\label{c2_1dspec.fig}
\end{figure}

\begin{figure}
\plotone{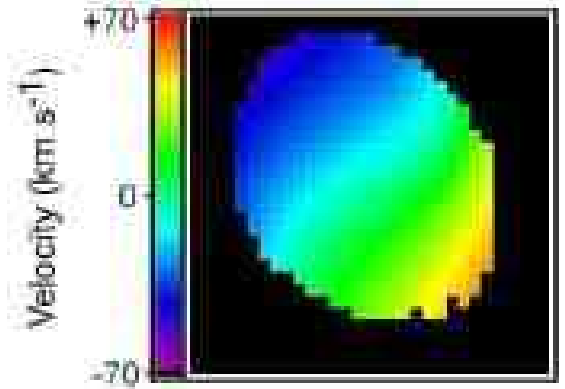}
\caption{Velocity map recovered from OSIRIS observations of DSF2237a-C2 after convolution with a 0.5'' PSF.  The field of view is identical to that
in Figure \ref{resultsfig.fig}.}
\label{c2blur.fig}
\end{figure}

For comparison to previous studies we produce an ``ideal'' long-slit spectrum of DSF2237a-C2 by 
extracting spaxels along the major axis (PA $38^{\circ}$, see \S 3) using a 180 mas wide ``slit'' and plot the resulting velocity
curve in Figure \ref{c2_1dspec.fig}.  The gradual trend of velocity across the galaxy is clear in this figure, along with the apparent
flattening of the velocity curve in the outskirts of the galaxy.
Although long-slit spectra were previously obtained with Keck-NIRSPEC observations
(Pettini et al. 2001), no shear was observed by
these authors despite close alignment of the NIRSPEC slit (PA 55$^{\circ}$) with the now-known shear axis (PA 38$^{\circ}$).  Given that the total
peak-to-peak amplitude of the shear ($\sim$ 140 km s$^{-1}$) should be identifiable with the spectral resolution of
NIRSPEC, we infer that DSF2237a-C2 represents  a practical case-example of how seeing-limited studies may miss
velocity structure on spatial scales smaller than that of the seeing halo.  
For comparison, in Figure \ref{c2blur.fig} we plot the 2-d velocity map recovered from our OSIRIS observations of DSF2237a-C2 after introducing
a 0.5'' spatial gaussian blur to roughly simulate the smearing effects of atmospheric seeing.
While the overall semblance of shear is retained, the peak-to-peak observed amplitude of this shear is reduced by a factor $\sim 2$.\footnote{We
note that Erb et al. (2004; their Fig. 3) reach a similar general conclusion based on observations of the galaxy Q1700-BX691, for which the apparent
amplitude of velocity shear varied by at least a factor of $\sim 2$ depending on the quality of seeing.}
In concert with the lower spectral resolution of NIRSPEC (in the low-resolution mode used for the Pettini et al. 2001 observations) and slight offsets
in slit placement and orientation from the optimal, it is easy to see why previous observations did not detect shear.

\begin{figure*}
\plotone{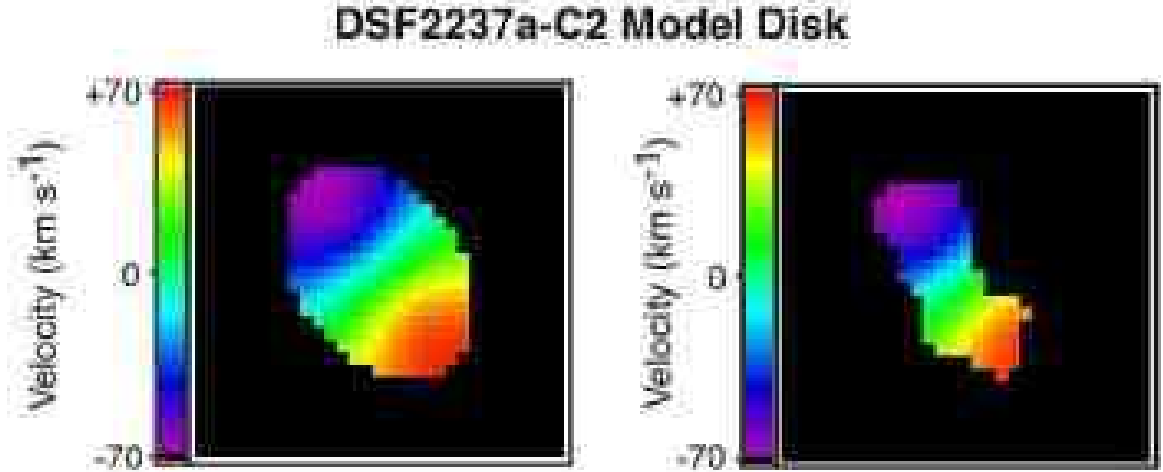}
\caption{{\it Left panel:} Idealized rotating disk model for DSF2237a-C2. {\it Right panel:} Velocity field which might be observed
with OSIRIS after convolution of the idealized disk model with the real PSF and selection of an appropriate distribution of bright ``emission'' regions.
Compare to the actual recovered velocity map shown in Figure \ref{resultsfig.fig} (bottom center panel). The field of view is identical to that
in Figure \ref{resultsfig.fig}.}
\label{c2_diskmodel.fig}
\end{figure*}

Adopting the inclination
and position angle derived in \S 3 from the nebular morphology, we create an idealized thin-disk model (i.e. a rotating sheet) which reproduces the
size and orientation of DSF2237a-C2 and adjust the asymptotic velocity and turnover radius as needed  to match the observations
shown in Figures \ref{resultsfig.fig} and \ref{c2_1dspec.fig}.
This model is characterized by a radius of 0.21''=1.6 kpc and a deprojected asymptotic velocity of 93 km s$^{-1}$,
implying a dynamical mass of  $3.2 \times 10^9 M_{\odot}$ interior to 1.6 kpc.
\footnote{This mass is slightly lower than that calculated on the basis of the velocity dispersion (and quoted in Table \ref{globals.table}),
reflecting in large part the uncertainty in the geometric prefactors used when converting from velocities to estimates of the dynamical mass.}
We plot the velocity map of this model disk in Figure \ref{c2_diskmodel.fig} (left panel), along with
the velocity map recovered from this model using simulated OSIRIS observations which include
convolution with the PSF and extraction of the high surface brightness pixels (right panel).
We have not attempted to match every detail of the observed velocity map by fitting a more complex model,
and some discrepancies remain (the blue-side velocities in particular are imperfectly matched by the simple model).
While more complicated fitting routines could certainly be performed (and indeed have been for galaxies studied by F{\"o}rster Schreiber et al. 2006 and other groups),
the exacting fits of such models frequently belie the large uncertainties inherent to the observations.
For example, the evidence for a turnover in the velocity curve of DSF2237a-C2 largely disappears with only slightly less careful
data reduction, and the salient features of the model are almost indistinguishable from similar models for which the inclination of the disk to the line
of sight differs by over 20$^{\circ}$.

Given the uncertainties in such quantitative models, what may we conclude about this galaxy with some confidence?  
It is unlikely that the shear in DSF2237a-C2 is due primarily to a major merger because the velocity field shows a smooth, gradual transition
rather than any sharp discontinuity as might be expected for two kinematically distinct components.  In addition, the flux distribution is smooth
and centrally concentrated; we do not detect multiple nuclei on scales more widely separated than $\sim$ 750 pc (the spatial resolution of the observation).
Adopting the simplest interpretation that the observed shear represents a rotating gaseous disk, it is worth asking whether
the disk is stable against gravitational collapse.
One traditional way of quantifying such stability is via the Toomre `$Q$' parameter (Toomre 1964) which effectively assesses the mass of the disk with respect
to the total gravitating mass of the system:
\begin{equation}
Q = \frac{V^2}{G M_{\rm disk}/r_{\rm disk}} = \frac{M_{\rm dyn}}{M_{\rm disk}}
\end{equation}
Typically, values of $Q < 1$ represent dynamically unstable systems for which the total baryonic mass exceeds that which can be supported by
the observed rotational velocity of the disk, indicating either that the disk itself is unstable or that non-rotational motions contribute a significant
degree of support to the system.  In contrast, values $Q > 1$ suggest that the observed baryonic component can be supported by the observed 
rotational motion alone, while $Q \sim 1$ represents the case of marginal stability.

Considering all of the uncertainties involved in determining the baryonic and dynamical masses, 
it is not possible to establish a single well-defined value for $Q$, although
realistic estimates for various cases place a limit $Q < 1$.  
In the first case, we consider the ratio of masses within the radius probed by the OSIRIS
detections ($r = 1.03$ kpc), for which $M_{\rm disk} = M_{\rm gas} + M_{\star} = 4.2 \times 10^{10} M_{\odot}$ in the limit that the entirety of the stellar mass
lies within this radius, and $M_{\rm disk} = 1.3 \times 10^{10} M_{\odot}$ in the opposite limit for which there is negligible stellar mass within
the central kpc.  Similarly, we may elect to use estimates of the dynamical mass based on the central velocity dispersion ($M_{\rm dyn} = 5 \times 10^{9} M_{\odot}$)
or the rotationally-supported thin disk model ($M_{\rm dyn} = 3.2 \times 10^9 M_{\odot}$).  Combining these assorted estimates, we obtain
a range of values $0.08 < Q < 0.38$.
In the second case, we calculate the global $Q$ value if we consider the mass within a much larger 10 kpc radius.
At such radii $M_{\rm disk} \geq 4.2 \times 10^{10} M_{\odot}$ (as the contribution of gas mass outside the central kiloparsec is unknown).  The dynamical mass
is even more uncertain\footnote{The dynamical mass may be particularly uncertain if the velocity curve continues rising out to larger radii instead of turning 
over at $\sim 93$ km s$^{-1}$ as suggested by observations.}: inserting $r = 10$ kpc in 
Equation \ref{dynmass.eqn} we find $M_{\rm dyn} = 5 \times 10^{10} M_{\odot}$, while extending the disk model
with asymptotic velocity 93 km s$^{-1}$ out to such radii gives $M_{\rm dyn} = 2 \times 10^{10} M_{\odot}$.
Adopting the highest physically motivated limit, we determine that $Q \leq 1.2$ within 10 kpc.
In both cases (and most noticeably within the central kpc), 
the baryonic mass in the system outweighs what could be supported by a simple rotating disk model, highlighting the importance of the
ionized gas velocity dispersion and indicating the improbability that star formation in DSF2237a-C2 occurs predominantly in
a rotationally supported thin disk of gas (see also discussion by F{\"o}rster Schreiber et al. 2006).
Interestingly, the kinematics of this galaxy are quite similar to those of the $z = 1.4781$ galaxy Q2343-BM133 observed by Wright et al. (2007),
which also had dispersion ($\sigma = 73 \pm 9$ km s$^{-1}$) comparable to its velocity shear (one-sided amplitude $v = 67$ km s$^{-1}$).

\section{DISCUSSION}

We have presented the results of a pilot study of the spatially resolved spectra of galaxies at redshift $z \sim 2 - 3$,
demonstrating that the velocity structure of the ionized gas appears to be inconsistent with
simple rotational support mechanisms.
While there has been some effort to target galaxies with particularly high star formation rates
to ensure detection in these early stages of the observational program, 
the H$\alpha$ flux of one of our three galaxies (Q0449-BX93, whose star formation rate was predicted from UV photometry)
is found to be quite close to the average (Erb et al. 2006a,b) of the $z \sim 2$ optically-selected galaxy sample.

It is, nevertheless, more than a little challenging to attempt to draw general conclusions about the entire $z \sim 2-3$ galaxy population
on the basis of only three galaxies, and therefore interesting to compare our results to integral-field velocity maps obtained recently by 
F{\"o}rster Schreiber et al. (2006) for 14 rest-UV selected galaxies at redshift $z \sim 2$
drawn from the catalog of Erb et al. (2006a) and
observed with SINFONI (Eisenhauer et al. 2003) on the VLT.  
While selection criteria varied, generally galaxies were known to be large and bright and (in many cases) known to exhibit considerable
velocity shear or dispersion on the basis of previous long-slit observations (Erb et al. 2003, 2006a).
Of these 14 galaxies, 9 show evidence of spatially resolved velocity gradients
with peak-to-peak amplitude ranging from 40 to 410 km s$^{-1}$ over $\sim$ 10 kpc.  Certainly, systems with strong kinematic
features appear to exist at $z \sim 2$, some of which may genuinely represent orbital velocities in massive rotating disks.
Indeed, the mean stellar mass of the subsample with shear greater than 50 km s$^{-1}$ 
is $9.4 \times 10^{10} M_{\odot}$ (based on the 6 galaxies for which $M_{\star}$ is given by Erb et al. 2006b), 
almost three times greater than the mean of the Erb et al. (2006b) sample
($<M_{\star}> = 3.6 \pm 0.4 \times 10^{10} M_{\odot}$).
In contrast, the remaining galaxies in the F{\"o}rster Schreiber et al. (2006) sample which do {\it not}
show such large-scale shear have mean stellar mass
$2.6 \times 10^{10} M_{\odot}$, similar to the masses of Q1623-BX453,
Q0449-BX93, and DSF2237a-C2 ($M_{\star} = 3.1,  \lesssim 1$, and $2.9 \times 10^{10} M_{\odot}$ respectively).
Rather than representing 
rare cases (such as counter-rotating mergers; F{\"o}rster Schreiber et al. 2006), 
the distribution of stellar masses suggests  that these non-rotating galaxies may instead be quite 
common in the $z \sim 2$ population.

Unfortunately,
it is difficult to directly compare the detailed kinematics of the F{\"o}rster Schreiber et al. (2006)
observations with our own since these data are seeing-limited
with PSF FWHM $\sim 0.6$'' (therefore smearing kpc-scale kinematics like those discussed in this contribution beyond recognizability; see
Fig. \ref{c2blur.fig}).
A more detailed comparison may be made to the $z = 2.3834$ galaxy BzK-15504 observed by Genzel et al. (2006) using SINFONI in combination
with natural guide-star adaptive optics for an angular resolution of 150 mas (i.e. comparable to the $\sim$ 110 - 150 mas resolution of our target
galaxies after smoothing).  While this galaxy exhibits resolved velocity shear in the outer regions of the galaxy
it is not well fit by a simple disk model within 3 kpc of the nucleus (i.e. approximately the region probed in our three target galaxies),
which Genzel et al. (2006) postulate may be due to radial gas flows fueling the central AGN whose presence is inferred
from rest-frame UV spectroscopy.  
One possibility may therefore be that disordered nuclear kinematics could generally join to more orderly shear at larger radii, although in our
data we find evidence for neither AGN nor large-radius shear.
Indeed, even in the circum-nuclear regions of BzK-15504 it is not obvious that a simple disk model provides an optimal description of the kinematics
since the line-of-sight velocity dispersion remains high at large radii
($\sigma \sim 60-100$ km s$^{-1}$, similar to the large values observed in DSF2237a-C2, and in Q2343-BM133 by Wright et al. 2007), 
and difference maps of the observed velocity field with the best-fitting
exponential disk model show deviations of greater than $100$ km s$^{-1}$ (see discussion by Wright et al. 2007).

What, then, may we conclude about the kinematics of these early galaxies and their implications
for galaxy formation?  One common explanation for both the unusual kinematics and the high nuclear gas fraction
is that these galaxies may represent major mergers of gas-rich galaxies.  While this explanation may suffice for a few cases however
(e.g. perhaps Q0449-BX93), it is unlikely (based on current cosmological models)
that this is the general explanation for the $z \sim 2$ population since the space density
of such galaxies with similar physical properties is far too high.  
Alternatively, efficient gas cooling in the early universe could have led to high cold gas fractions in early-stage
galactic disks which then fragmented due to self-gravity and collapsed to form a nuclear starburst (e.g. Immeli et al. 2004a).
In both pictures, considerable kinematic structure should exist in these galaxies although
(similar to local ULIRGS; Colina et al. 2005) the kinematics of the brightest nebular emission regions
may be relatively featureless and quite variable in detail from system to system.
However, it is unknown whether the lack of shear (or indeed any significant velocity structure beyond a high dispersion)
persists to lower surface-brightness ionized gas at larger radii or if the kinematics of this gas are strong and disordered,
ordered with dynamically cold orbital motion, or if indeed
significant nebular line emission exists at all beyond
the radii probed in this study.  For one of the galaxies studied here (Q1623-BX453) previous deep long-slit
spectroscopy (Erb et al. 2006b) could have detected any significant shear if it were present at larger radii
(and the kinematic major axis was reasonably well-aligned with the slit), but found no evidence for such a trend.

What is perhaps most clear is that we do not yet understand the dynamical state of galaxies during
this period in which they are forming the bulk of their stars.  
While traditional theories of galaxy formation posit that the majority of star formation should occur in rotationally supported
gaseous disks, this does not appear to be in good agreement with the accumulating observational evidence,
inviting the construction of new models of gas cooling which can explain the flat, dispersion-dominated kinematic features
of galaxies in the young universe.


\acknowledgements
We would like to thank Andrew Benson and Juna Kollmeier for numerous helpful discussions,
and Naveen Reddy for providing {\it Spitzer} photometry for Q1623-BX453.
The authors also thank Randy Campbell, Al Conrad, David LeMignant, and Jim Lyke
for their invaluable help obtaining the observations presented herein.
D.R.L. and C.C.S. have been supported by grants AST-0606912
and AST-0307263 from the US National Science Foundation.
Finally, we wish to extend thanks to those of Hawaiian ancestry on whose sacred mountain
we are privileged to be guests.

\end{document}